\documentclass[printer]{aa}
\usepackage{natbib}
\usepackage{aas_macros}
\usepackage{epstopdf}

\usepackage{graphicx}
\usepackage[T1]{fontenc}
\usepackage{amsmath}
\usepackage{amssymb}
\usepackage{amsmath}
\usepackage{graphicx}
\usepackage{amsmath}
\usepackage{amssymb}
\usepackage{amsfonts}
\usepackage{graphicx,epsfig,psfig,graphics}
\usepackage{aas_macros}
\usepackage{graphics}
\usepackage{color}

\def\be{\begin{eqnarray}}
\def\ee{\end{eqnarray}}
\graphicspath{{}}

\titlerunning{CMB reconstruction from the WMAP and Planck PR2 data}
\authorrunning{Bobin et al.}

\title{CMB reconstruction from the WMAP and Planck PR2 data}
\author{ \hspace{0.25in} J. Bobin\inst{1}, F. Sureau\inst{1}, J.-L. Starck\inst{1}}
\institute{$^1$ Laboratoire AIM, UMR CEA-CNRS-Paris 7, Irfu, SAp/SEDI, Service d'Astrophysique, CEA Saclay, F-91191 GIF-SUR-YVETTE CEDEX, France.}

\begin{document}
 

\abstract{In this article, we describe a new estimate of the Cosmic Microwave Background (CMB) intensity map reconstructed by a joint analysis of the full Planck 2015 data (PR2) and WMAP nine-years. It provides more than a mere update of the CMB map introduced in \citep{PR1_LGMCA} since it benefits from an improvement of the  component separation method L-GMCA (Local-Generalized Morphological Component Analysis) that allows the efficient separation of correlated components \citep{AMCA15}. Based on the most recent CMB data, we further confirm previous results \citep{PR1_LGMCA} showing that the proposed CMB map estimate exhibits appealing characteristics for astrophysical and cosmological applications: i) it is a full sky map that did not require any inpainting or interpolation post-processing, ii) foreground contamination is showed to be very low even on the galactic center, iii) it does not exhibit any detectable trace of thermal SZ contamination. We show that its power spectrum is in good agreement with the Planck PR2 official theoretical best-fit power spectrum. Finally, following the principle of reproducible research, we provide the codes to reproduce the L-GMCA, which makes it the only reproducible CMB map.}

\date{Received -; accepted -}
\maketitle
%

\section{Introduction}
The Cosmic Microwave Background (CMB) provides a snapshot of our Universe at the time of recombination. It is therefore a unique probe for the cosmologists since it carries crucial information about the dawn of the Universe as well as its evolution to the current state. The ability to estimate a clean full-sky estimation of the CMB map is critical to accurately study its statistical properties, which depend on the primordial fluctuations from which they arose.\\
Maps of the CMB have been reconstructed from the frequency channels of the WMAP mission \citep{WMAP9_1}, the Planck mission in 2013 \citep{PR1_compsep}. In 2015, the Planck consortium released the full Planck data (Planck Release 2 or PR2), from which CMB maps have been derived \citep{PR2_compsep}. Thanks to a longer time of integration, the instrumental noise has been significantly reduced with respect to the Planck PR1 data released in 2013. Data calibration has been dramatically improved, which further reduces the impact of instrumental systematics \citep{PR2_Overview}. Understanding the beam transfer function \citep{PR2_LFI, PR2_HFI} largely reduces the uncertainty of the data, especially at small scales.\\
The reconstruction of a clean CMB map from multiple frequency channels is well known to be a challenging task since it requires removing spurious astrophysical foreground contaminations. These emissions mainly include:
\begin{itemize}
\item{\it In the low frequency regime, below $100$GHz:} The predominant source of contamination comes from the galactic synchrotron and free-free emissions \citep{Gold_WMAP7Templates,PR2_frg}, especially at large angular scale. The spinning dust \citep{PER_SDust,PR2_LFG} highly correlates with the thermal dust emission.\\
\item{\it In the high frequency regime, above $100$GHz:} At higher frequencies, the dominant source of contamination is the thermal dust emission \citep{PER_Dust}, especially at the vicinity of the galactic center. The Cosmic Infrared Background \citep{PR1_CIB} provides a significant emission at all galactic latitudes at the highest frequencies ({\it i.e.} $545$ and $857$GHz).\\
\end{itemize}
These galactic foregrounds mainly limit our ability to accurately estimate the CMB map at the vicinity of the galactic center, where these emissions become largely dominant. The presence of instrumental systematics and noise further hinders the estimation of the CMB map.\\
Estimating a clean CMB map from frequency channels has been showed to be best carried out by component separation methods \citep{Leach_08,PR1_compsep,PR2_compsep}. To that respect, the CMB maps released by the Planck consortium in 2015 have been computed using four different component separation techniques: Commander \citep{Commander}, Needlet ILC (NILC - \citep{NeedletILC}), SEVEM \citep{WSEVEM} and SMICA \citep{ica:Del2003}. These methods have been extensively decribed in \citep{PR2_compsep}.

In \citep{PR1_LGMCA}, we introduced a new CMB map using a recent sparsity-based component separation technique, coined Local-Generalized Morphological Component Analysis (L-GMCA - see \citep{2012arXiv1206.1773B}). The proposed CMB map was further based on a joint processing of the Planck 2013 (PR1) and WMAP nine-year data. We showed that, in constrast to the available CMB maps, the proposed L-GMCA-based estimate is a clean full-sky estimate that did not require any masking or inpainting technique for large scale statistical studies \citep{PR1_LGMCA_anomalies}. Furthermore, we demonstrated that the L-GMCA CMB map was the only one to explicitly project out the thermal SZ effect \citep{PR1_LGMCA}. For these reasons, the L-GMCA CMB map has been used for different cosmological studies \citep{BenDavid14,Aiola14,Lanusse14,2015JCAP...09..011L,2015JCAP...06..047N}.

\subsection*{Contributions}

This paper presents a novel estimation of the Cosmic Microwave Background map reconstructed from the Planck 2015 data (PR2) and the WMAP nine-year data \citep{WMAP9_1}, which first updates the CMB map we published in \citep{PR1_LGMCA}. This new map is based on the sparse component separation method L-GMCA (Local-Generalized Morphological Component Analysis - see \citep{2012arXiv1206.1773B}). Additionally, it benefits from the latest advances in this field \citep{AMCA15}, which allows to accurately discriminate between correlated components. In the sequel, we show that this new map presents significant improvements with respect to the available CMB map estimates:
\begin{itemize} 
\item It is a full-sky map, based on the latest CMB data, that did not require any interpolation or inpainting technique.
\item It provides a very clean estimation of the galactic region with very low level of foreground residuals.
\item It has no detectable thermal SZ contamination
\end{itemize}
Furthermore, its power spectrum is in good agreement with the Planck 2015 best fit $C_\ell$ \citep{PR2_PS} on $76 \%$ of the sky.\\
The rest of the paper will be organized as follows: Section \ref{sec:compsep} briefly describes the improved L-GMCA method and its application to the Planck PR2 and WMAP nine-year data. Evaluations and comparisons of the L-GMCA CMB map with the official Planck 2015 CMB maps are detailed in section~\ref{sec:results}.

\section{Sparse component separation for CMB reconstruction}
\label{sec:compsep}
\paragraph*{The L-GMCA algorithm.}
From the viewpoint of applied mathematics, reconstructing the CMB from multi-frequency data is known as a Blind Source Separation problem (see \citep{JutBook} for a general review). In this framework, the observations are assumed to be a linear combination of $n$ components so that the $m$ frequency channels verify at each pixel $k$ :
\begin{equation}
\forall i=1,\cdots,m; \, x_i[k] = \sum_{j=1}^n a_{ij} s_j[k] + z_i[k]
\end{equation}
where $s_j$ stands for the $j$-th component, $a_{ij}$ is a scalar that models for the contribution of the $j$-th component to channel $i$ and $z_i$ models the instrumental noise or model imperfections. This problem is customarily recast into the matrix formulation :
\begin{equation}
{\bf X} = {\bf A S} + {\bf Z}
\end{equation}
The ability to accurately discriminate between the components to be reconstructed highly depends on the precise modeling of their statistical properties. With the exception of Commander, which is a parameterized Bayesian method, the CMB maps that were derived for processing the Planck PR2 data all build upon second-order statistics.\\
In contrast, the L-GMCA method (Local-Generalized Morphological Component Analysis - \citep{2012arXiv1206.1773B}) is based on a different separation principle: the sparse modeling of the components. Such a modeling is motivated by the fact that foreground components have a sparse distribution in the wavelet domain: few wavelet coefficients contain most of the energy of the components. It is very important to recall that modeling the data in the wavelet domain only changes its statistical distribution without altering its information content. Sparse signal modeling in the wavelet domain is very well adapted to efficiently describe the statistics of non-Gaussian and non-stationary processes. This has been showed to significant improve the extraction of foreground components such as galactic foregrounds and point sources \citep{2012arXiv1206.1773B,Sureau14}.\\
In the sequel, we will define $\boldsymbol{A}$ as the mixing matrix and $\boldsymbol{\Phi}$ as a wavelet transform, we assume that each source $s_{j}$ can be sparsely represented in ${ \bf \Phi}$; $s_{j}=\alpha_{j}\boldsymbol{\Phi}$, where $\mathbf{\alpha}$ is a $N_{s}\times T$ matrix whose rows are $\alpha_{j}$.
The multichannel noiseless data $\boldsymbol{Y}$ can be written as
\begin{equation}
\boldsymbol{Y}=\boldsymbol{A}\mathbf{\alpha}\boldsymbol{\Phi}\:.\label{eq:tensor1-1}
\end{equation}
The L-GMCA algorithm estimates mixing parameters ({\it i.e.} the mixing matrix $\bf A$ and the components $\bf S$) which yields the sparsest components $\boldsymbol{S}$. This is formalized by the following optimization problem :
\begin{equation}
\label{eq:estim_gmca}
\min\frac{1}{2}\left\Vert \boldsymbol{X}-\boldsymbol{A}\mathbf{\alpha}\boldsymbol{\Phi}\right\Vert _{F}^{2}+\lambda \left \| \mathbf{\alpha} \right \|_{p}^{p}\:,
\end{equation}
where typically $p=0$
and ${\bf \left\Vert \boldsymbol{X}\right\Vert }_{\mathrm{F}}= \sqrt { \left(\textrm{trace}(\boldsymbol{X}^{T}\boldsymbol{X})\right)}$
is the Frobenius norm.\\ \\

Similarly to the version used to analyse the Planck PR1 data, the properties of the L-GMCA algorithm we used to process the WMAP nine-year and Planck 2015 data are described as follows: i) it accurately accounts for the individual point-spread function (PSF) of each individual frequency channel, ii) it implements a mechanism to deal with the spatial variations of the components' spectra, and iii) a sparsity-based post-processing technique is used to provide a clean estimation of the galactic center. Full details are given in \citep{PR1_LGMCA}.

\paragraph*{Improvement of the L-GMCA algorithm.}

Additionally, the new version of the L-GMCA algorithm includes very recent advances in Blind Source Separation. Indeed, most component separation methods are customarily based on some separation principle to discriminate between the components to be estimated. This includes their statistical independence or their decorrelation. However, most foreground components in full-sky microwave observations do not verify these basic statistical assumptions: i) at small and medium scales ({\it i.e.} typically for $\ell > 100$), distinct components such as synchrotron and thermal dust emissions have high emissivity in similar regions of the sky, especially at the vicinity of the galactic center. As well, components such as spinning dust and thermal dust emissions are well-known to be spatially correlated. These types of similarities lead to the presence of partial correlations between components with different physical nature. At large-scale ({\it i.e.} typically for $\ell < 100$), only few spherical harmonics are observed. Subsequently, even theoretically decorrelated components exhibit some correlation, customarily named chance-correlations, which can be regarded as partial correlation between the components. However, it has been well established that partial correlations dramatically hamper the efficiency of component separation methods \citep{AMCA14,AMCA15}.\\ 

To limit the impact of these partial correlations, the current implementation of the L-GMCA algorithm includes very recent advances in blind source separation \citep{AMCA15}. This refinement relies on the fact that partial correlations between components will very likely impact a small subset of the sparse wavelet coefficients of the components. This algorithm then builds upon a weighting scheme that penalizes correlated entries, which are the most detrimental for separation. For that purpose, the problem described in Equation~\ref{eq:estim_gmca} is substituted with the following minimization problem :
\begin{equation}
\label{eq:estim_amca}
\min_{{\bf A},{\bf S}}\frac{1}{2}\mbox{Trace}\left\{ \left(\boldsymbol{X}-\boldsymbol{A}\mathbf{\alpha}\boldsymbol{\Phi}\right) {\bf \Phi}^T{\bf \Pi}{\bf \Phi} \left(\boldsymbol{X}-\boldsymbol{A}\mathbf{\alpha}\boldsymbol{\Phi}\right)^T\right\}+\lambda \left \| \mathbf{\alpha} \right \|_{p}^{p}\:,
\end{equation}
where the matrix $\bf \Pi$ is a diagonal weight matrix. According to \citep{AMCA15}, partial correlations between the components are very likely to be traced by non-sparse columns of the source matrix $\bf S$ in the wavelet domain, which is given by ${\bf S}{\bf \Phi}^T$. Therefore, the weighting strategy we proposed in \citep{AMCA15} consists in penalizing samples of the sources with respect to their sparsity level in the wavelet domain using the aforementioned weighting scheme. More precisely, given an estimate of the components $\hat{\bf S}$, the weighting matrix will be computed as follows :
\begin{equation}
\forall i=1,\cdots,T; \quad \left[{\bf \Pi}\right]_{i,i} = \frac{1}{\|[\hat{\bf S}{\bf \Phi}^T]^{(i)}\|_q + \epsilon}
\end{equation}
where $\left[{\bf \Pi}\right]_{i,i}$ is the $i$-th diagonal element of $\bf \Pi$, $[\hat{\bf S}{\bf \Phi}^T]^{(i)}$ is the $i$-th column of $\hat{\bf S}{\bf \Phi}^T$, $\epsilon$ is a small scalar to avoid numerical issues, and $q \leq 1$ so as to measure the sparsity level of the components. Based on this relationship, the weight matrix $\bf \Pi$ is updated iteratively with respect to the estimated sources during the minimization process. For an exhaustive description of this extension, we refer the reader to \citep{AMCA15}.\\

\paragraph*{Application of the L-GMCA to the WMAP nine-year and Planck PR2 data}

The L-GMCA algorithm is precisely defined by a set of parameters, which are similar to the one we used to analyze the Planck PR1 data in \citep{PR1_LGMCA}. The wavelet transform is defined by filters in the spherical harmonic domain, which are identical to the bands described in \citep{PR1_LGMCA}. Similarly to \citep{PR1_LGMCA}, the IRIS map \citep{IRIS05} has been added as an extra observation. Its use is limited to a region that is located on the galactic center. For more details about the application of the L-GMCA, we refer the interested reader to \citep{2012arXiv1206.1773B} and \citep{PR1_LGMCA}.\\

In the remaining of this paper the CMB map derived from the joint analysis of the Planck PR2 and WMAP9 data will be denoted by WPR2 L-GMCA.

\section*{Reproducible research}

\begin{table*}[htbp]
  \centering
  \begin{tabular}{@{} lcl @{}} 
  Product name  &Type & Description\\
  \hline
  Planck WPR2 products:\\
  WPR2\_CMB\_muK\_hr1.fits   &  Map  & WPR2 CMB estimate, first half ring  \\
  WPR2\_CMB\_muK\_hr2.fits   &  Map  & WPR2 CMB estimate, second half ring\\
  WPR2\_CMB\_muK.fits   &  Map  & WPR2 CMB map estimate \\
  WPR2\_CMB\_noise\_muK.fits   &  Map  & WPR2 noise map estimate \\
  \hline 
  Software products:\\
  {\tt run\_L-GMCA\_wpr2\_getmaps.pro}  & code (IDL)  & code to compute the CMB map estimates\\ & & (requires {\tt HealPix} and {\tt iSAP}). \\
  {\tt wpr2\_analysis\_routines.pro}  & code (IDL)  & routines to reproduce the figures of the paper. \\ & & (requires {\tt HealPix} and {\tt iSAP}).\\
  \hline

  \end{tabular}
  \caption{List of products made available in this paper in the spirit of reproducible research, available here: \url{http://www.cosmostat.org/research/cmb/planck_wpr2/}.}
  \label{tab:reproducible_ps}
\end{table*}
In the spirit of participating in reproducible research, we have made all codes and resulting products that constitute the main results of this paper public. In Table \ref{tab:reproducible_ps} we list all the products that will be made freely available as a result of this paper and which will be available. The L-GMCA code, the L-GMCA CMB maps as well as the CMB power spectrum will be made available (\footnotesize{\url{http://www.cosmostat.org/research/cmb/planck_wpr2/}}).


\section{Conclusion}
We provide a new CMB map based on a combination of the WMAP nine-year and Planck PR2 data using sparse component analysis. It builds upon the L-GMCA algorithm, sparsity-based component separation method, with the addition of the most recent advances on sparse component separation.  We show that the proposed estimation procedure yields a clean full-sky CMB map with no significant foreground residuals on the galactic center. In contrast to the currently available CMB maps derived from the Planck 2015 data, the proposed L-GMCA WPR2 map contains no detectable tSZ contamination. The properties of the L-GMCA WPR2 map can be summarized as follows:
\begin{itemize}
\item it is the only one full-sky clean CMB map based on the Planck PR2 data that does not require any inpainting techniques.
\item it is free of tSZ contamination, which makes it the unique reasonable candidate for kSZ studies.
\item  its power spectrum using a $76\% $ mask is in good agreement with the Planck best fit theoretical power spectrum.
\end{itemize}
Finally, on the contrary to most other CMB maps, the L-GMCA WPR2  is also fully reproducible, which is fundamental in order to well control 
all systematics in any further statistical study such as non-Gaussianity detection or CMB lensing.


\begin{acknowledgements}
This work was funded by the PHySIS project (contract no. 640174) and the DEDALE project (contract no. 665044), within the H2020 Framework Program of the European Commission. We used Healpix software \citep{Healpix}, iSAP\footnote{\url{http://www.cosmostat.org/software/isap/}} software, \emph{WMAP} data \footnote{\url{http://map.gsfc.nasa.gov}},  and Planck data  \footnote{\url{http://map.gsfc.nasa.gov}} .
\end{acknowledgements}

\bibliographystyle{aa} 
\bibliography{gmca_bib}

\begin{thebibliography}{35}
\expandafter\ifx\csname natexlab\endcsname\relax\def\natexlab#1{#1}\fi

\bibitem[{{Abrial} {et~al.}(2007){Abrial}, {Moudden}, {Starck}, {Bobin},
  {Fadili}, {Afeyan}, \& {Nguyen}}]{inpainting:abrial06}
{Abrial}, P., {Moudden}, Y., {Starck}, J., {et~al.} 2007, jfaa, 13, 729--748

\bibitem[{Aiola {et~al.}(2014)Aiola, Kosowsky, \& Wang}]{Aiola14}
Aiola, S., Kosowsky, A., \& Wang, B. 2014, ArXiv 1410.6138

\bibitem[{Ben-David \& Kovetz(2014)}]{BenDavid14}
Ben-David, A. \& Kovetz, E. 2014, MNRAS, 445, 2116

\bibitem[{{Bennett}(2013)}]{WMAP9_1}
{Bennett}, C.~L., e. 2013, ApJS

\bibitem[{{Bobin} {et~al.}(2014{\natexlab{a}}){Bobin}, Rapin, {Starck}, \&
  Larue}]{AMCA14}
{Bobin}, J., Rapin, J., {Starck}, J.-L., \& Larue, A. 2014{\natexlab{a}}, in
  Proceedings of IEEE ICIP

\bibitem[{{Bobin} {et~al.}(2015){Bobin}, Rapin, {Starck}, \& Larue}]{AMCA15}
{Bobin}, J., Rapin, J., {Starck}, J.-L., \& Larue, A. 2015, IEEE Transactions
  on Signal Processing, 63, 1199

\bibitem[{{Bobin} {et~al.}(2013){Bobin}, {Starck}, {Sureau}, \&
  {Basak}}]{2012arXiv1206.1773B}
{Bobin}, J., {Starck}, J.-L., {Sureau}, F., \& {Basak}, S. 2013, Astronomy \&
  Astrophysics, 550, A73

\bibitem[{{Bobin} {et~al.}(2014{\natexlab{b}}){Bobin}, {Sureau}, {Starck},
  {Rassat}, \& {Paykari}}]{PR1_LGMCA}
{Bobin}, J., {Sureau}, F., {Starck}, J.-L., {Rassat}, A., \& {Paykari}, P.
  2014{\natexlab{b}}, A\&A, 563

\bibitem[{Comon \& Jutten(2010)}]{JutBook}
Comon, P. \& Jutten, C. 2010, Handbook of blind source separation (Elsevier)

\bibitem[{Delabrouille {et~al.}(2009)Delabrouille, Cardoso, Jeune, Betoule,
  Fay, \& Guilloux}]{NeedletILC}
Delabrouille, J., Cardoso, J.-F., Jeune, M.~L., {et~al.} 2009, Astronomy and
  Astrophysics, 493, 835

\bibitem[{Delabrouille {et~al.}(2003)Delabrouille, Cardoso, \&
  Patanchon}]{ica:Del2003}
Delabrouille, J., Cardoso, J.-F., \& Patanchon, G. 2003, Monthly Notices of the
  Royal Astronomical Society, 346, 1089-1102

\bibitem[{{Eriksen} {et~al.}(2008){Eriksen}, {Jewell}, {Dickinson}, {Banday},
  {G{\'o}rski}, \& {Lawrence}}]{Commander}
{Eriksen}, H.~K., {Jewell}, J.~B., {Dickinson}, C., {et~al.} 2008, APJ, 676, 10

\bibitem[{{Fern{\'a}ndez-Cobos} {et~al.}(2012){Fern{\'a}ndez-Cobos}, {Vielva},
  {Barreiro}, \& {Mart{\'{\i}}nez-Gonz{\'a}lez}}]{WSEVEM}
{Fern{\'a}ndez-Cobos}, R., {Vielva}, P., {Barreiro}, R.~B., \&
  {Mart{\'{\i}}nez-Gonz{\'a}lez}, E. 2012, \mnras, 420, 2162

\bibitem[{{Gold} {et~al.}(2011){Gold}, {Odegard}, {Weiland}, {Hill}, {Kogut},
  {Bennett}, {Hinshaw}, {Chen}, {Dunkley}, {Halpern}, {Jarosik}, {Komatsu},
  {Larson}, {Limon}, {Meyer}, {Nolta}, {Page}, {Smith}, {Spergel}, {Tucker},
  {Wollack}, \& {Wright}}]{Gold_WMAP7Templates}
{Gold}, B., {Odegard}, N., {Weiland}, J.~L., {et~al.} 2011, APJS, 192, 15

\bibitem[{{Gorski} {et~al.}(2005){Gorski}, Hivon, Banday, Wandelt, Hansen,
  Reinecke, \& {Bartelmann}}]{Healpix}
{Gorski}, K., Hivon, E., Banday, A.~J., {et~al.} 2005, ApJ, 622

\bibitem[{{Hivon} {et~al.}(2002){Hivon}, {G{\'o}rski}, {Netterfield}, {Crill},
  {Prunet}, \& {Hansen}}]{2002ApJ...567....2H}
{Hivon}, E., {G{\'o}rski}, K.~M., {Netterfield}, C.~B., {et~al.} 2002, ApJ,
  567, 2

\bibitem[{Lanusse {et~al.}(2014)Lanusse, {Paykari}, {Starck}, {Sureau},
  {Bobin}, \& {Rassat}}]{Lanusse14}
Lanusse, F., {Paykari}, P., {Starck}, J., {et~al.} 2014, Astronomy \&
  Astrophysics, 571

\bibitem[{Leach(2008)}]{Leach_08}
Leach, S. 2008, Astronomy \& Astrophysics, 491

\bibitem[{{Luzzi} {et~al.}(2015){Luzzi}, {G{\'e}nova-Santos}, {Martins}, {De
  Petris}, \& {Lamagna}}]{2015JCAP...09..011L}
{Luzzi}, G., {G{\'e}nova-Santos}, R.~T., {Martins}, C.~J.~A.~P., {De Petris},
  M., \& {Lamagna}, L. 2015, \jcap, 9, 11

\bibitem[{{Miville-Desch{\^e}nes} \& {Lagache}(2005)}]{IRIS05}
{Miville-Desch{\^e}nes}, M.-A. \& {Lagache}, G. 2005, APJSS, 157

\bibitem[{{Notari} \& {Quartin}(2015)}]{2015JCAP...06..047N}
{Notari}, A. \& {Quartin}, M. 2015, \jcap, 6, 47

\bibitem[{{Planck Collaboration}(2014)}]{PR1_CIB}
{Planck Collaboration}. 2014, A\&A, 571

\bibitem[{{Planck Collaboration}(2015{\natexlab{a}})}]{PR2_Overview}
{Planck Collaboration}. 2015{\natexlab{a}}, A\&A

\bibitem[{{Planck Collaboration}(2015{\natexlab{b}})}]{PR2_compsep}
{Planck Collaboration}. 2015{\natexlab{b}}, A\&A

\bibitem[{{Planck Collaboration}(2015{\natexlab{c}})}]{PR2_LFI}
{Planck Collaboration}. 2015{\natexlab{c}}, A\&A

\bibitem[{{Planck Collaboration}(2015{\natexlab{d}})}]{PR2_HFI}
{Planck Collaboration}. 2015{\natexlab{d}}, A\&A

\bibitem[{{Planck Collaboration}(2015{\natexlab{e}})}]{PR2_frg}
{Planck Collaboration}. 2015{\natexlab{e}}, Astronomy \& Astrophysics

\bibitem[{{Planck Collaboration}(2015{\natexlab{f}})}]{PR2_PS}
{Planck Collaboration}. 2015{\natexlab{f}}, A\&A

\bibitem[{{Planck Collaboration}(2015{\natexlab{g}})}]{PR2_LFG}
{Planck Collaboration}. 2015{\natexlab{g}}, A\&A, ArXiv:1506.06660

\bibitem[{{Planck Collaboration} {et~al.}(2011{\natexlab{a}}){Planck
  Collaboration}, {Ade}, {Aghanim}, {Armitage-Caplan}, {Arnaud}, {Ashdown},
  {Atrio-Barandela}, {Aumont}, {Baccigalupi}, {Banday}, \& et~al.}]{PER_Dust}
{Planck Collaboration}, {Ade}, P.~A.~R., {Aghanim}, N., {et~al.}
  2011{\natexlab{a}}, A\&A, 536, A19

\bibitem[{{Planck Collaboration} {et~al.}(2011{\natexlab{b}}){Planck
  Collaboration}, {Ade}, {Aghanim}, {Armitage-Caplan}, {Arnaud}, {Ashdown},
  {Atrio-Barandela}, {Aumont}, {Baccigalupi}, {Banday}, \& et~al.}]{PER_SDust}
{Planck Collaboration}, {Ade}, P.~A.~R., {Aghanim}, N., {et~al.}
  2011{\natexlab{b}}, A\&A, 536, A20

\bibitem[{{Planck Collaboration} {et~al.}(2013{\natexlab{a}}){Planck
  Collaboration}, {Ade}, {Aghanim}, {Armitage-Caplan}, {Arnaud}, {Ashdown},
  {Atrio-Barandela}, {Aumont}, {Baccigalupi}, {Banday}, \&
  et~al.}]{PR1_compsep}
{Planck Collaboration}, {Ade}, P.~A.~R., {Aghanim}, N., {et~al.}
  2013{\natexlab{a}}, ArXiv e-prints 1303.5072

\bibitem[{{Planck Collaboration} {et~al.}(2013{\natexlab{b}}){Planck
  Collaboration}, {Ade}, {Aghanim}, {Arnaud}, {Ashdown}, {Aumont},
  {Baccigalupi}, {Baker}, {Balbi}, {Banday}, \& et~al.}]{PR1_PS}
{Planck Collaboration}, {Ade}, P.~A.~R., {Aghanim}, N., {et~al.}
  2013{\natexlab{b}}, ArXiv e-prints 1303.5075

\bibitem[{{Rassat} {et~al.}(2014){Rassat}, {Starck, J.-L.}, {Paykari},
  {Sureau}, \& {Bobin}}]{PR1_LGMCA_anomalies}
{Rassat}, A., {Starck, J.-L.}, {Paykari}, P., {Sureau}, F., \& {Bobin}, J.
  2014, Journal of Cosmology and Astroparticle, 08

\bibitem[{{Sureau} {et~al.}(2014){Sureau}, {Starck}, {Bobin}, {Paykari}, \&
  {Rassat}}]{Sureau14}
{Sureau}, F., {Starck}, J., {Bobin}, J., {Paykari}, P., \& {Rassat}, A. 2014,
  Astronomy \& Astrophysics, 566

\end{thebibliography}

\clearpage

\begin{appendix}
\label{sec:results}

\section{Map and Power Spectrum Estimation}


\begin{figure}[htb]
\centerline{
\hbox{\includegraphics [scale=0.2]{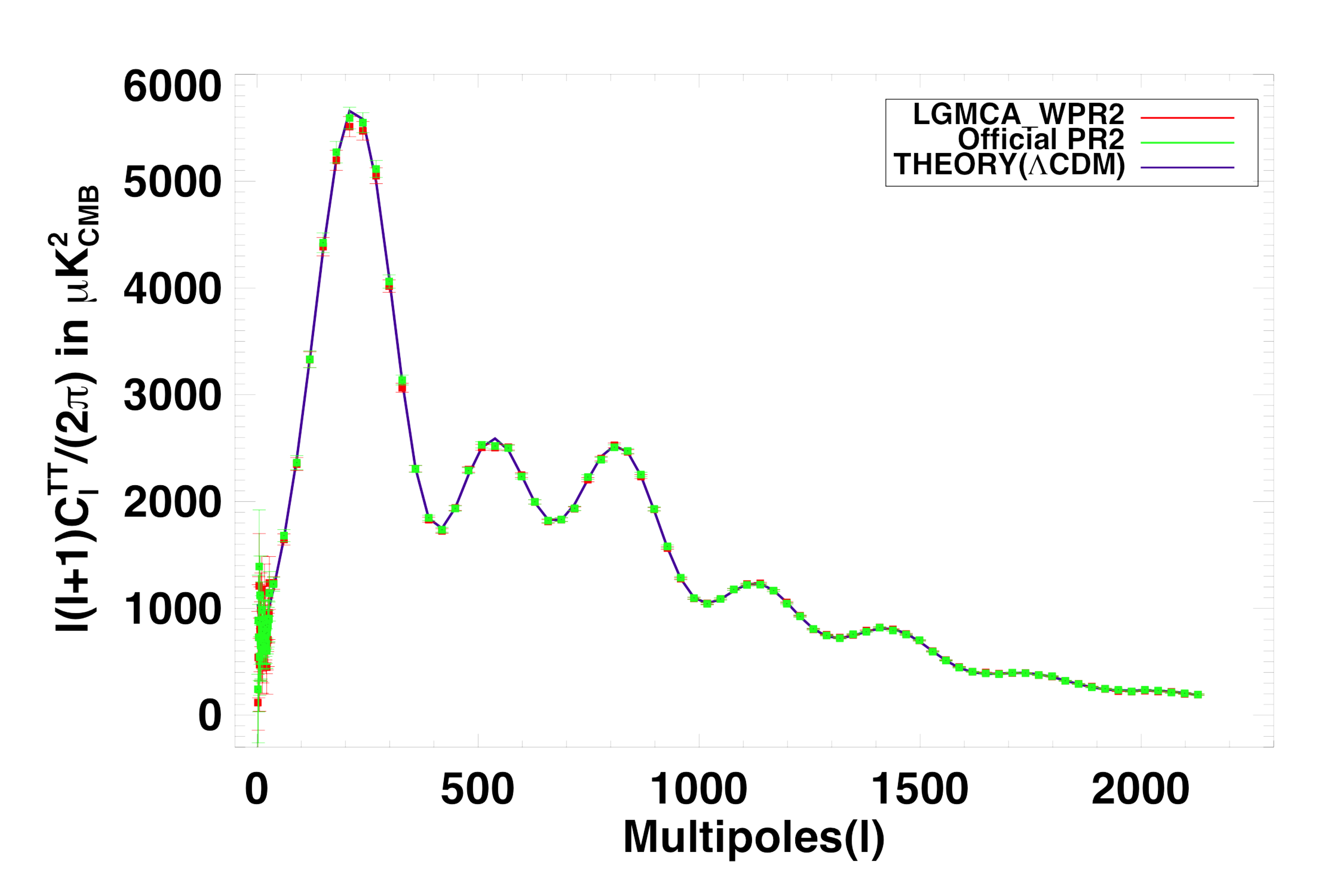}
}
}
\centerline{
\hbox{\includegraphics [scale=0.2]{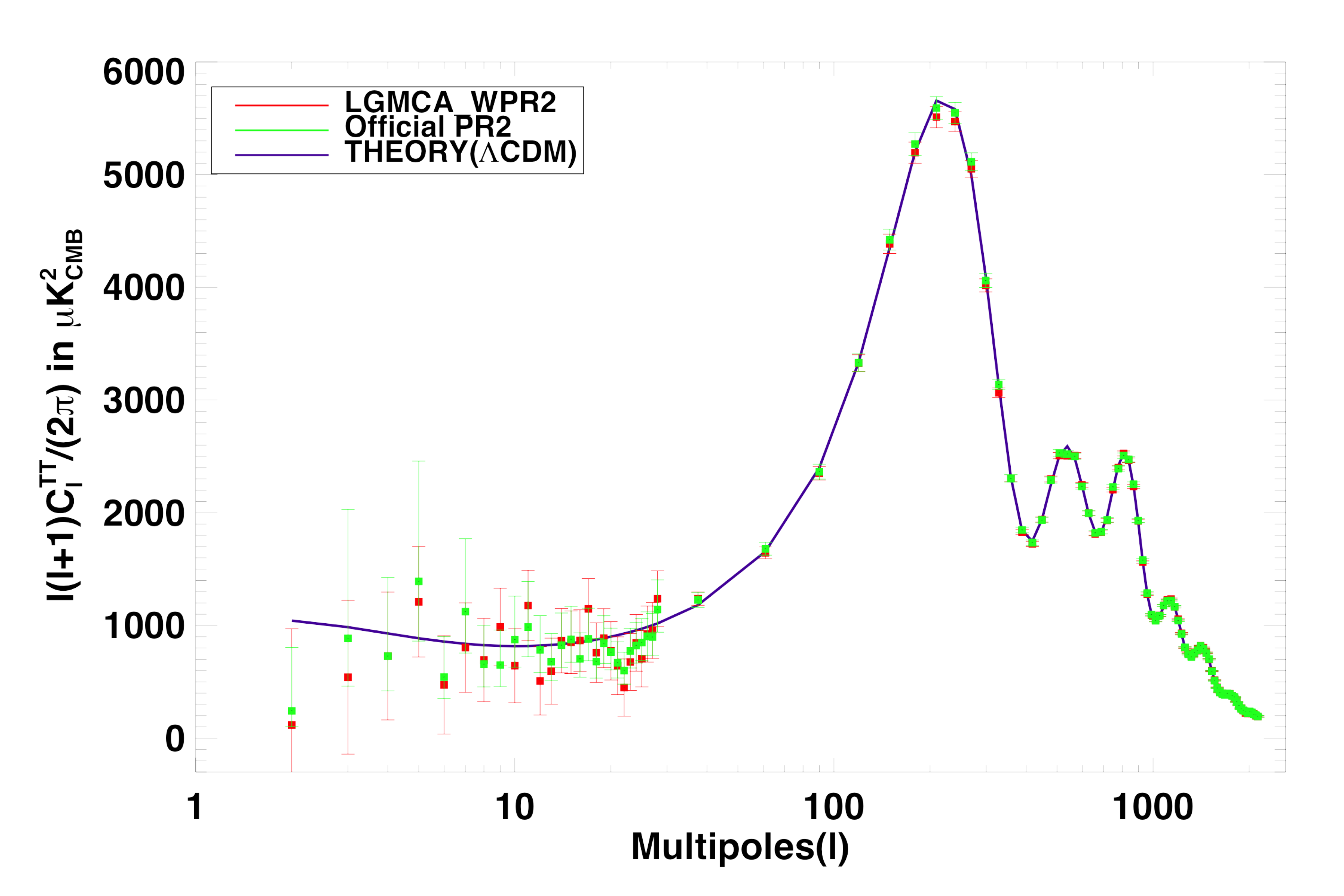}
}
}
\caption{Top, estimated power spectrum of the WPR1 L-GMCA map (red). The solid black line is the Planck-only best-fit $C_\ell$ provided by the Planck consortium. Bottom, power spectrum in logarithmic scale. Error bars are set to $1\, \sigma$.}
\label{fig_ps}
\end{figure}

\begin{figure}[htb]
\centerline{
\hbox{\includegraphics [scale=0.2]{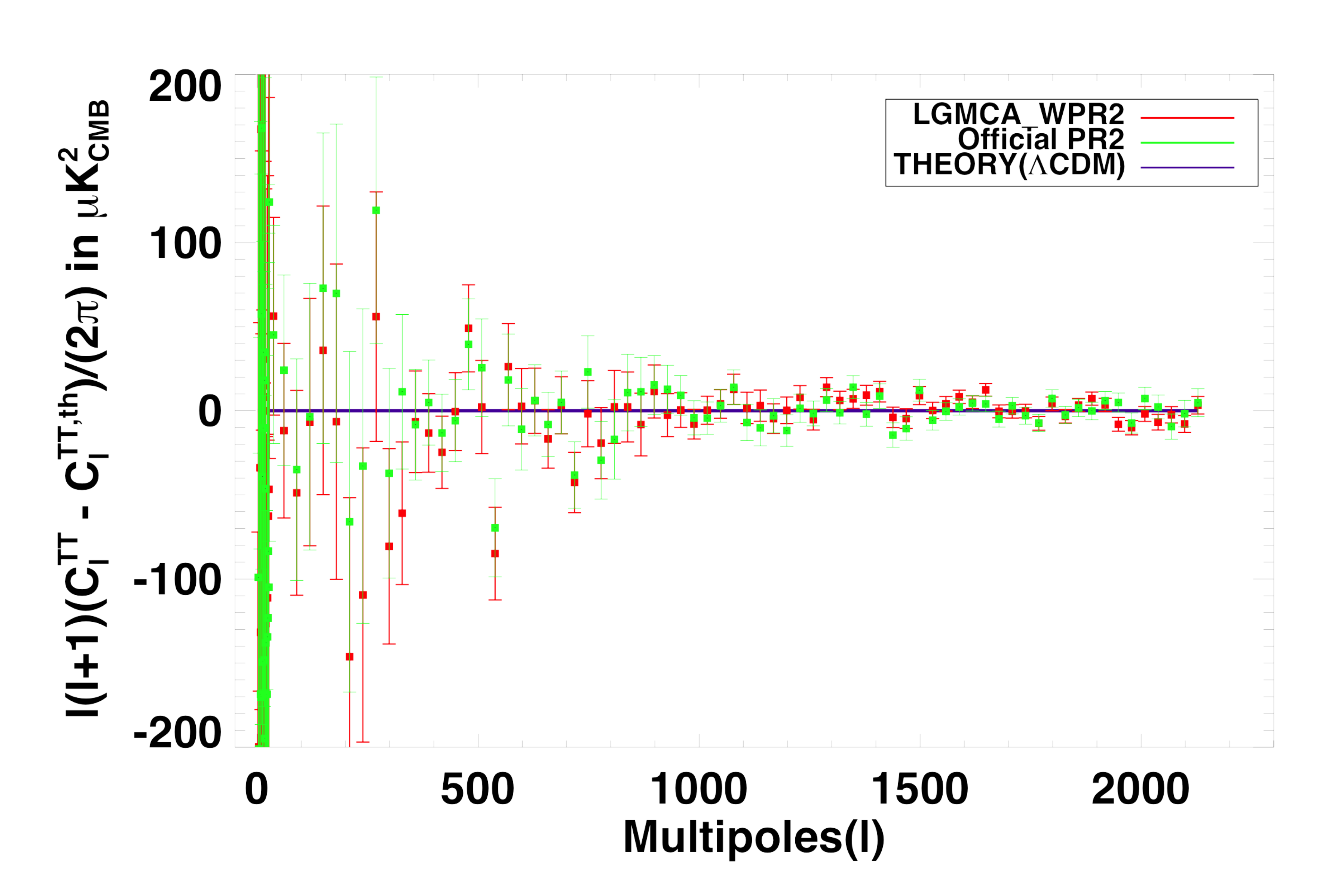}
}
}
\centerline{
\hbox{\includegraphics [scale=0.2]{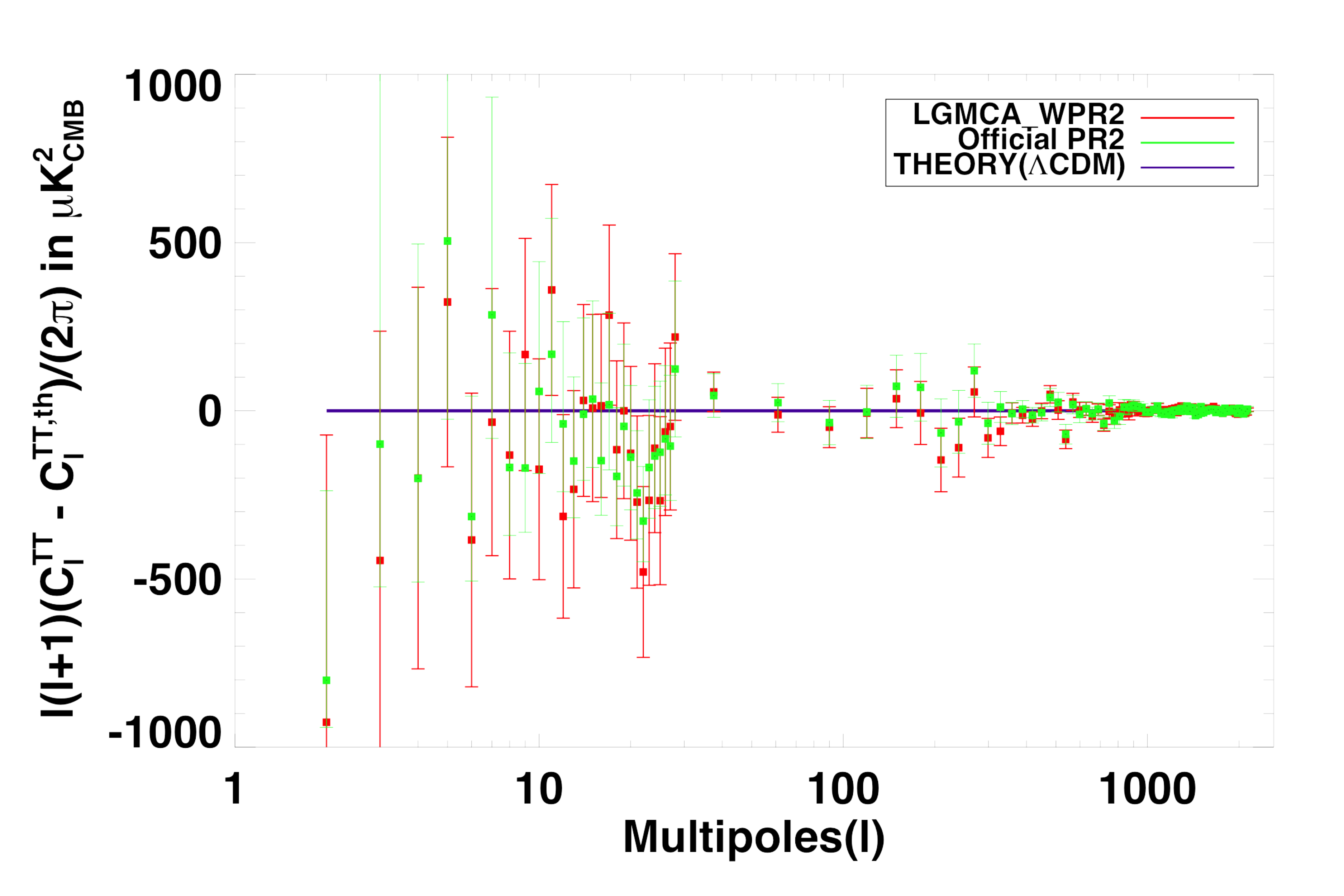}
}
}
\caption{Top, difference between the power spectrum estimated from the WPR1 L-GMCA map and the Planck-only best-fit $C_\ell$ provided by the Planck consortium. Bottom, difference between the estimated and theoretical power spectra in logarithmic scale. Error bars are set to $1\, \sigma$.}
\label{fig_resi_ps}
\end{figure}

Following \citep{PR1_LGMCA}, the L-GMCA algorithm is applied to the five WMAP nine-year maps (offset-corrected maps) and the nine averages of the Planck PR2 half ring maps. The CMB is then derived by applying the inverse of the mixing matrices estimated with the L-GMCA algorithm. Estimates of the noise maps are derived from random noise realizations for the WMAP channels. These realizations have been computed using the noise covariance matrices provided by the WMAP consortium. To estimate noise maps from the Planck PR2 data, half differences of half mission maps have been used as a proxy for a single data noise realization.\\
Next in this article, the CMB power spectrum is estimated by computing the cross-correlation between the two half-ring maps. In contrast to the CMB signal, noise decorrelates between half rings; the noise bias then vanishes when cross-correlating half-ring maps. Subsequently, the cross-correlation provides an estimate of the CMB power spectrum that is free of any noise-related bias. In the case of WMAP, such virtual half rings maps can be obtained by calculating the difference and sum of the WMAP data and a single data noise realization.\\
The power spectrum is evaluated from a sky coverage of $76\%$; the corresponding common mask is composed of a point source and galactic mask chosen from the Planck consortium masks, which is identical to the analysis mask used in \citep{PR1_LGMCA}.\\

The estimation procedure is described by the following processing steps:
\begin{enumerate}
\item{\it Inpainting the masked maps:} Following \citep{PR1_LGMCA}, the masked maps are first inpainted prior to deconvolution to avoid potential processing artifacts especially at the vicinity of remaining point sources. This procedure keeps the estimation of the power spectrum unaltered since it is eventually evaluated from the $76\%$ of the sky which are kept unchanged through the inpainting step.\\
\item{\it Deconvolution of the maps:} the masked and inpainted maps are deconvolved to infinite resolution up to $\ell = 3200$.\\
\item{\it Mask correction:} Correcting for the effect of the mask is made by applying the MASTER mask deconvolution technique \citep{2002ApJ...567....2H}.\\
\item{\it Correcting for unresolved point sources:} As noticed in \citep{PR1_LGMCA}, the CMB power spectrum is still biased by the contamination of the unresolved point sources, especially at high $\ell > 1500$. Following \citep{PR1_PS,PR1_LGMCA}, a first-order correction for the contribution of the unresolved point sources is a constant or scale-independent power spectrum. We apply exactly in the sequel the same correction to estimate the CMB power spectrum.
\end{enumerate}
The CMB power spectrum we estimated from the L-GMCA map is displayed in Figure~\ref{fig_ps}. The error bars account for the cosmic variance and the noise-related variance. These plots show that the the power spectrum derived from the WPR2 L-GMCA map is in good agreement with the official Planck 2015 best-fit theoretical power spectrum in the range $\ell \in [0,2000]$. 


\section{Evaluation of the L-GMCA CMB map}


\begin{figure*}[htb]
\centerline{
\hbox{
\includegraphics [scale=0.25]{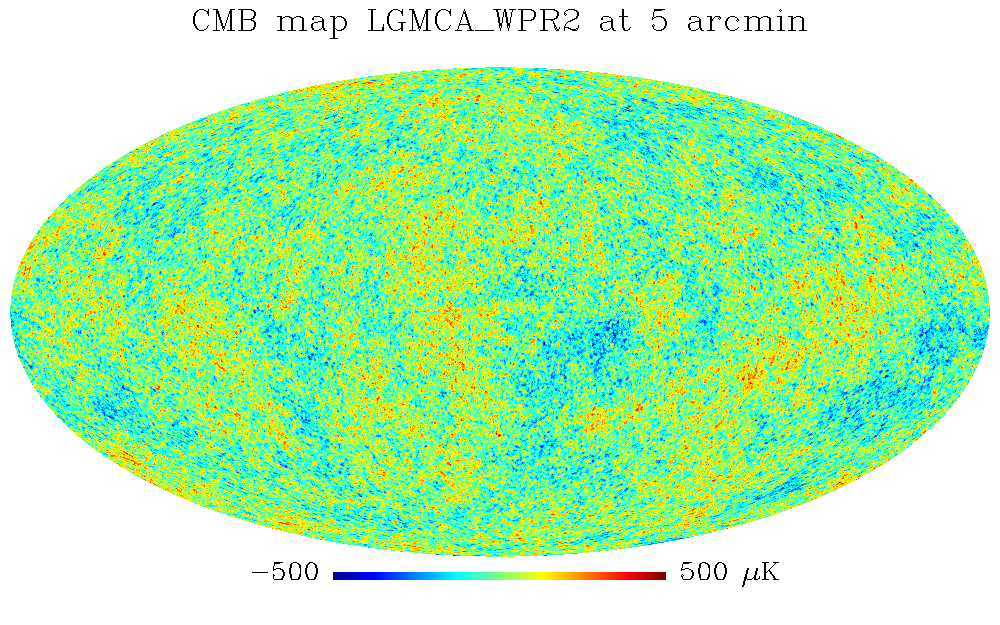}
}
}
\caption{CMB map estimated using the L-GMCA algorithm from the WMAP nine-year and Planck PR2 data.}
\label{fig:CMBmaps}
\end{figure*}

\begin{figure*}[htb]
\centerline{
\hbox{
\includegraphics [scale=0.3]{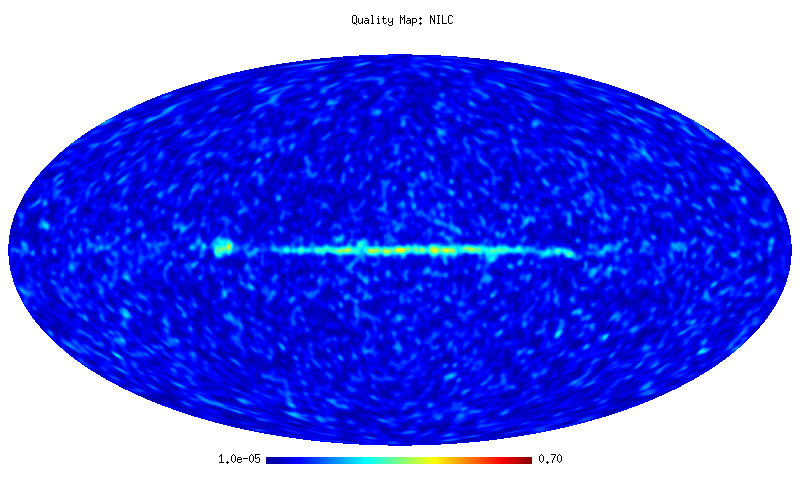}
\includegraphics [scale=0.3]{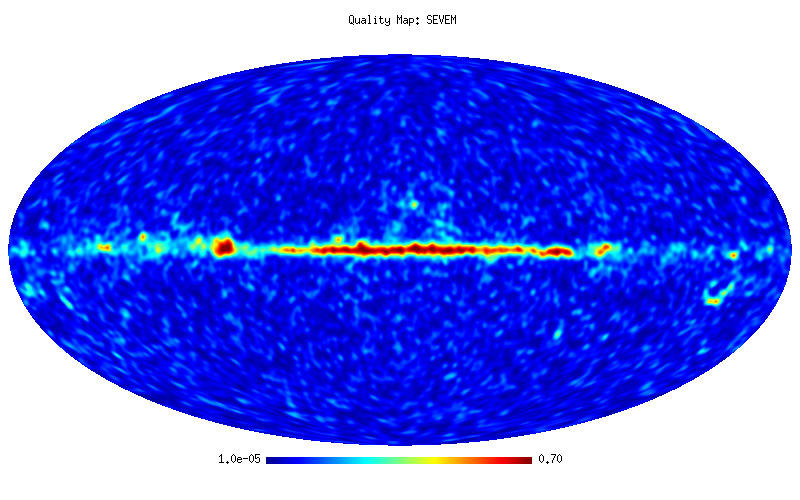}
}}
\centerline{
\hbox{
\includegraphics [scale=0.3]{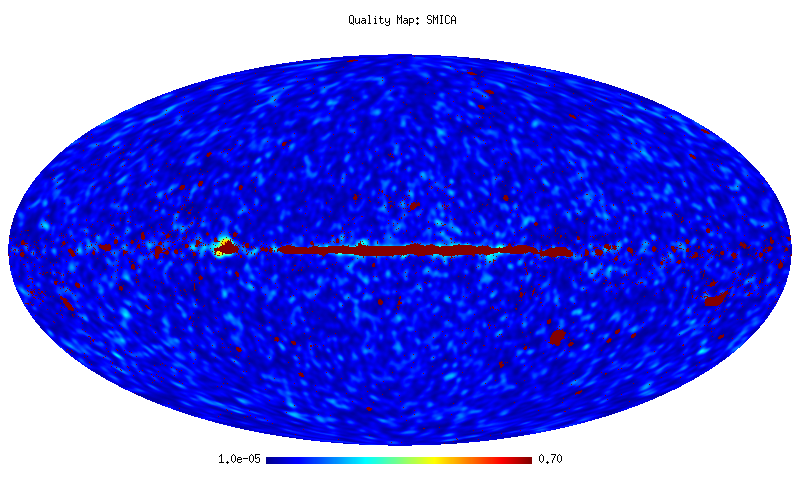}
\includegraphics [scale=0.3]{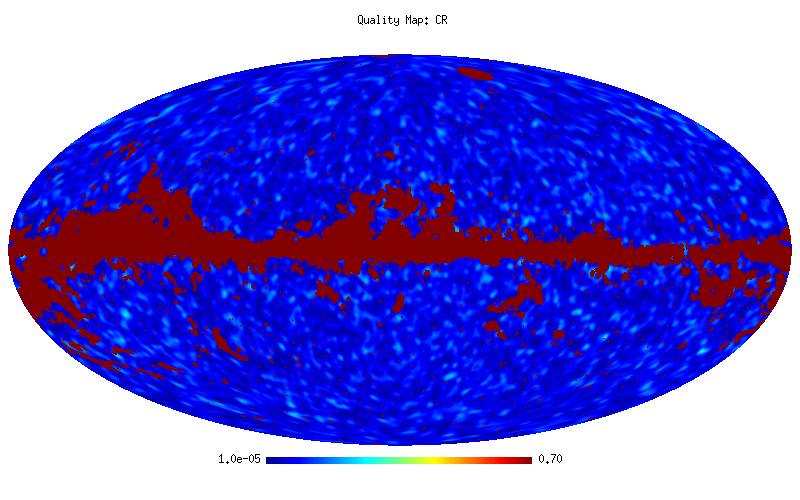}
}
}
\centerline{
\hbox{
\includegraphics [scale=0.3]{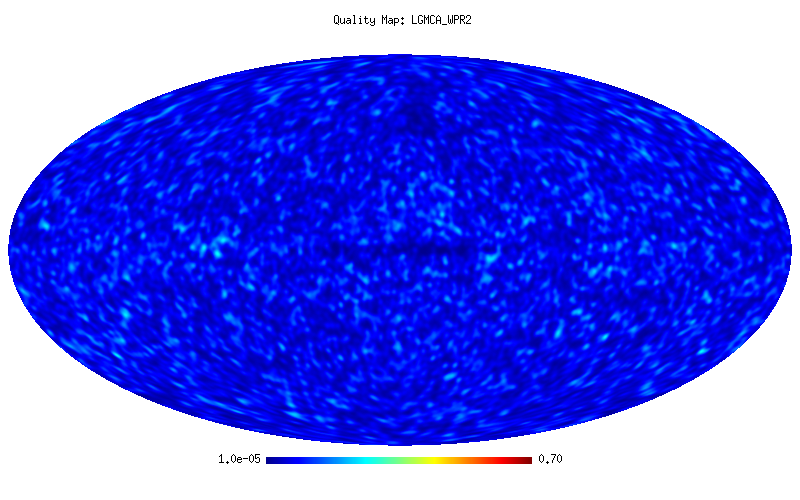}
}
}
\caption{Top, PR2 NILC and SEVEM CMB quality maps. Middle, PR2 SMICA and PR2 Commander  quality maps. Bottom, WPR2 L-GMCA quality map.}
\label{fig_qualmap}
\end{figure*}

Following \citep{PR1_LGMCA}, we assessed various measures of contamination signatures to evaluate potential deviations from the expected characteristics of the CMB map and perform comparisons with the four official CMB maps provided by the Planck consortium in 2015.

\subsection{The quality map: a measure of the CMB map power excess}
In \citep{PR1_LGMCA}, we emphasized that estimating the quality of a reconstructed CMB map from real data without any strong assumption about the expected map is a challenging task. To that purpose, we defined the so-called {\it quality map}, which only relies on the assumption that the $\Lambda$-CDM best fit power spectrum $C_\ell$ provides a good approximation of the expected power of the CMB in the spherical harmonic domain. Therefore, this allows to compare the local deviation around each pixel $k$ in the estimated CMB map to its expected power that the best-fit $C_\ell$ indicates and assess its compatibility with the expected noise level. We refer to \citep{PR1_LGMCA} for a precise definition of the quality map.\\

Let us recall that the SMICA and Commander maps have been inpainted; they therefore have a partial sky coverage. As a consequence, we set to zero the pixels in $Q$ which turn to be inpainted. The quality maps are displayed in Figure~\ref{fig_qualmap}. In this figure significant excess are traced by yellow and red colors. Dark blue pixels are likely related to statistical fluctuations. Similarly to \citep{PR1_LGMCA}, the SEVEM and NILC maps exhibit more contamination in the galactic plane than SMICA and L-GMCA. Outside the galactic plane, none of these maps present significant contamination.

 
\subsection{Galactic center estimation}

\begin{figure*}[htb]
\centerline{
\hbox{
\includegraphics [scale=0.35]{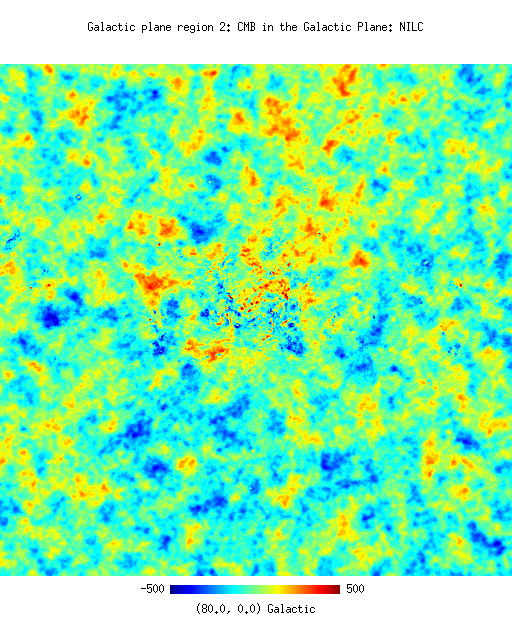}
\includegraphics [scale=0.35]{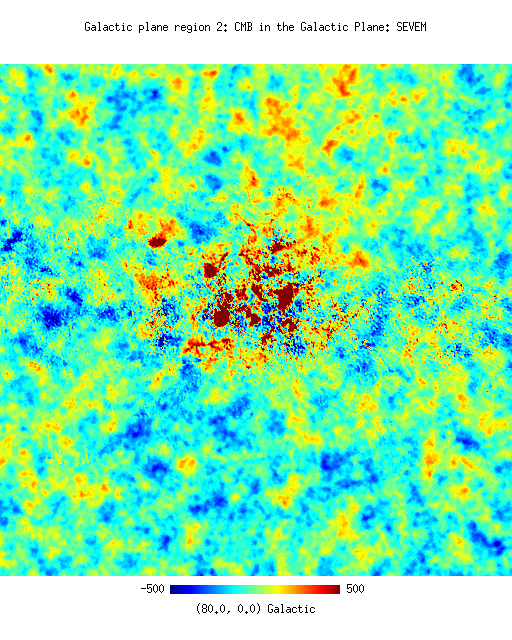}
}}
\centerline{
\hbox{
\includegraphics [scale=0.35]{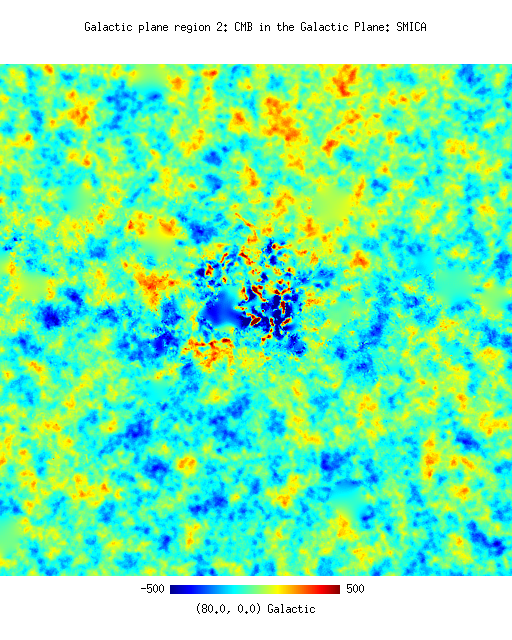}
\includegraphics [scale=0.35]{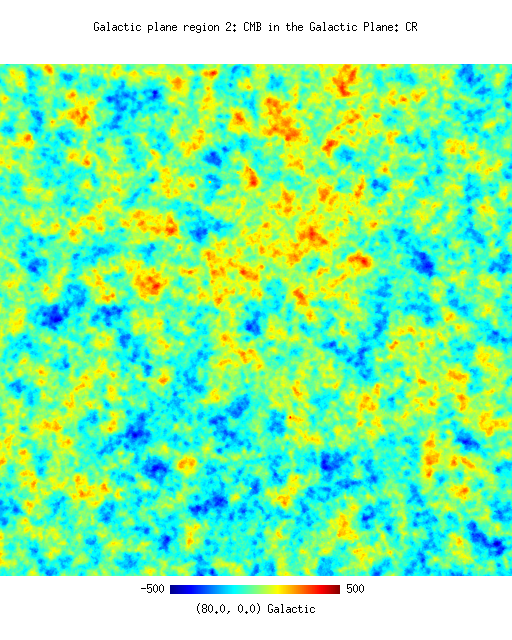}
}
}
\centerline{
\hbox{
\includegraphics [scale=0.35]{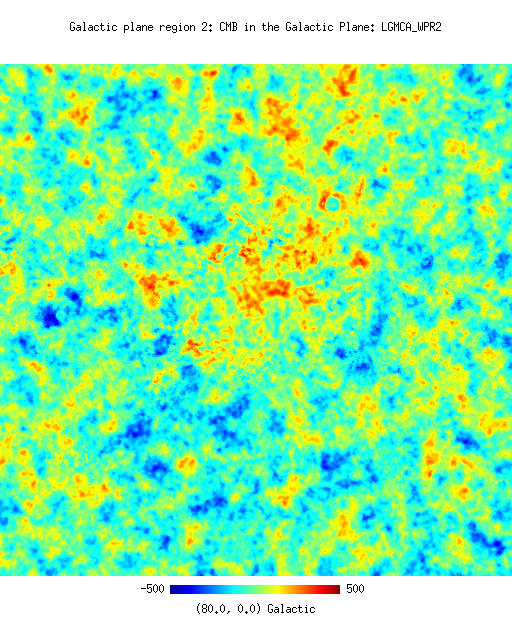}
}
}
\caption{galactic center region, {centered at} (l,b)=(80.0,0). Top, PR2  NILC and SEVEM CMB maps, and bottom, PR2 SMICA and Commander maps and WPR2 L-GMCA CMB maps.}
\label{fig_galcenter}
\end{figure*}


Figure~\ref{fig:CMBmaps} reveals that foreground residuals are mainly visible at the vicinity of the galactic plane. To better highlight these differences, we display in the Figure~\ref{fig_galcenter} an illustrative region that is located in the galactic center. Let us notice that, in this region, the Commander and SMICA maps have been post-processed using an inpainting technique. In the case of the Commander map, a significant portion of the sky (up to $18 \%$ for $\ell > 1000$) has been filled in with a constrained Gaussian realization. With the exception of the Commander map, the maps released by the Planck consortium all exhibit significant foreground residuals. Conversely, the L-GMCA map does not show any visible foreground emission residual. Similarly to the WPR1 L-GMCA we published in \citep{PR1_LGMCA}, the joint processing of the WMAP and Planck data allows a better separation of the foreground components, which further leads to a cleaner estimate of the CMB map in the galactic center.

\subsection{Detecting traces of SZ Contamination}

\begin{figure*}[htb]
\centerline{
\hbox{
\includegraphics [scale=0.3]{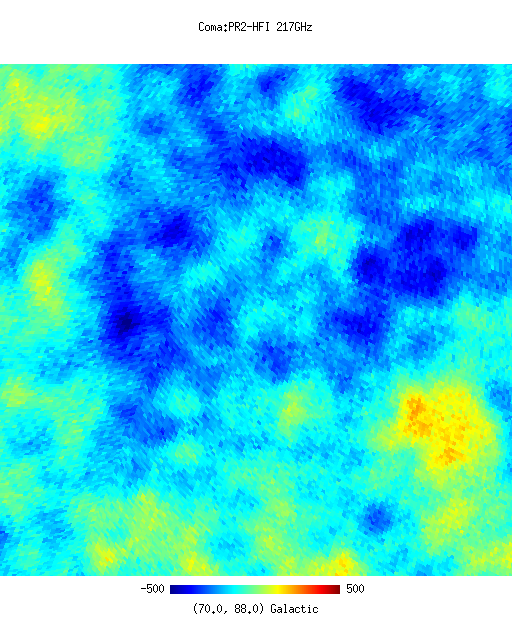}
\includegraphics [scale=0.3]{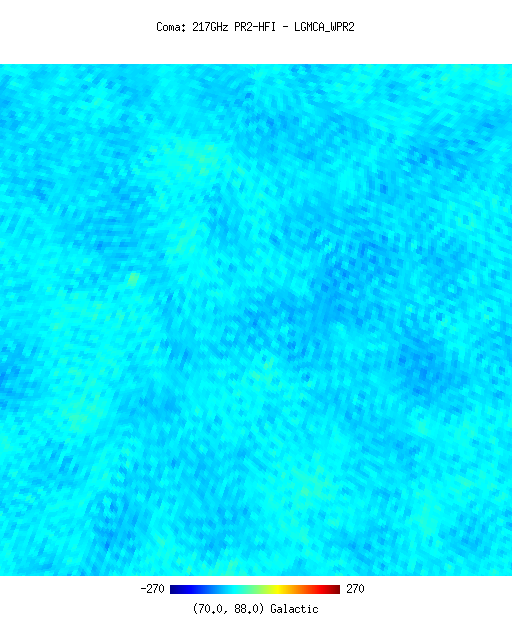}
}}
\centerline{
\hbox{
\includegraphics [scale=0.3]{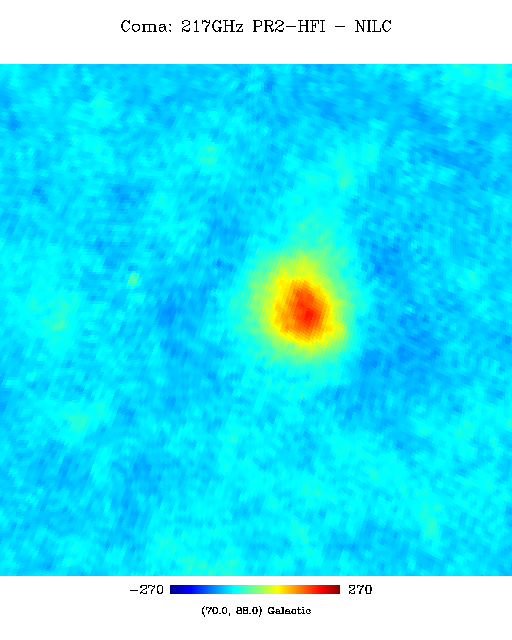}
\includegraphics [scale=0.3]{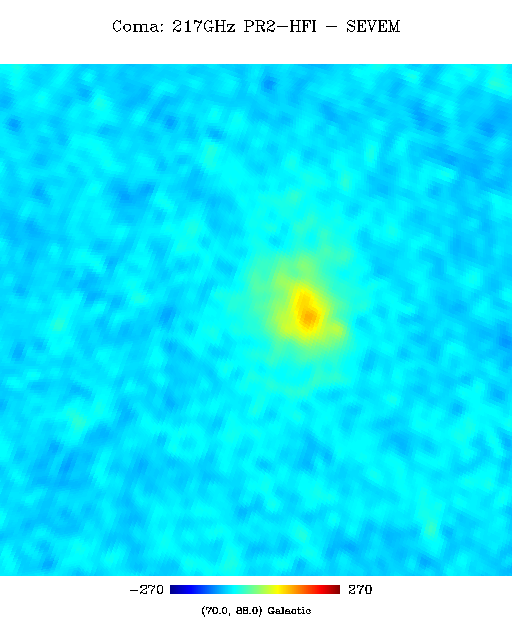}
}}
\centerline{
\hbox{
\includegraphics [scale=0.3]{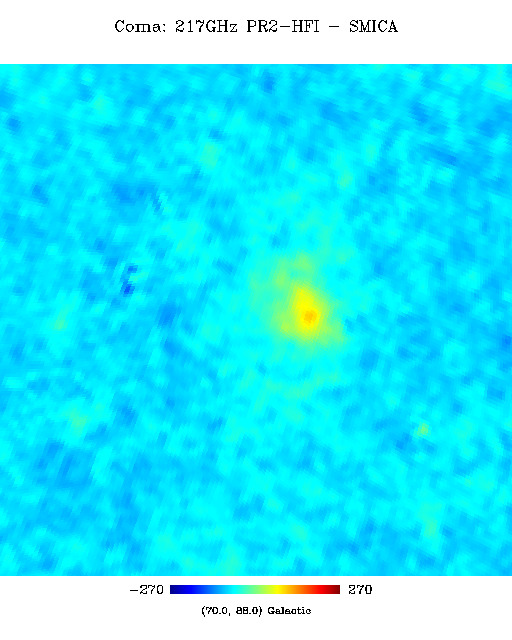}
\includegraphics [scale=0.3]{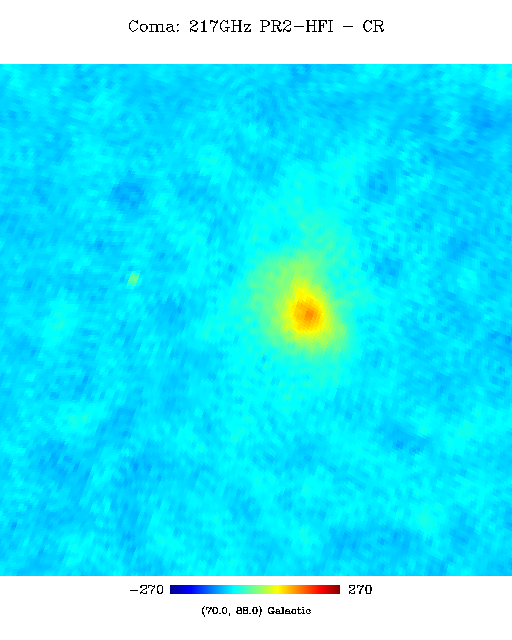}
}
}
\caption{Coma cluster area. Top, HFI-217 GHz map and  difference map between 
HFI-217GHz and WPR2 L-GMCA CMB map. Middle and bottom, difference map between 
HFI-217GHz and CMB maps, respectively PR2 NILC, SEVEM, SMICA and Commander.}
\label{fig_coma}
\end{figure*}

In \citep{PR1_LGMCA}, we noticed that another major difference between the CMB maps released by the Planck consortium and the L-GMCA CMB map was that the latter did not exhibit any detectable trace of thermal Sunyaev Zel'Dovich (tSZ) effect. For that purpose, we made profit of an interesting property of the tSZ effect: its contribution can be neglected at $217$GHz. Subsequently, the difference between the HFI-$217$GHz channel map 
and any estimate of the CMB map should cancel out the CMB without revealing tSZ residuals.\\
Figure \ref{fig_coma} shows the difference maps of the estimated CMB maps with the PR2 HFI frequency map at $217$GHz. The Coma cluster appears clearly in the four maps provided by the Planck consortium. Conversely, the L-GMCA map does not exhibit any detectable trace of tSZ contamination, since
L-GMCA method is the only one to explicitly project out the thermal SZ emission during the component separation. This therefore makes the proposed L-GMCA CMB map the perfect candidate to study the kinetic SZ effect.

\subsection{Higher-order statistics of the reconstructed CMB maps}

An effective evaluation of the level of foreground contamination can be performed by evaluating the Gaussianity or non-Gaussianity of the estimated CMB map. To that purpose and as advocated in \citep{PR1_LGMCA}, higher-order statistics (HoS) provide a model-independent measure of non-Gaussianity (NG).\\ 
One major difficulty is that all the maps have not been post-processed in the same way: some of them have been inpainted. To perform fair comparisons, we chose to first inpaint all the maps with a combination of our point source mask and the SMICA mask of inpainted pixels (retaining $95\%$ of the sky) using the sparse inpainting technique described in \citep{inpainting:abrial06}.\\

\begin{figure*}[htb]
\centerline{
\hbox{
\includegraphics [scale=0.25]{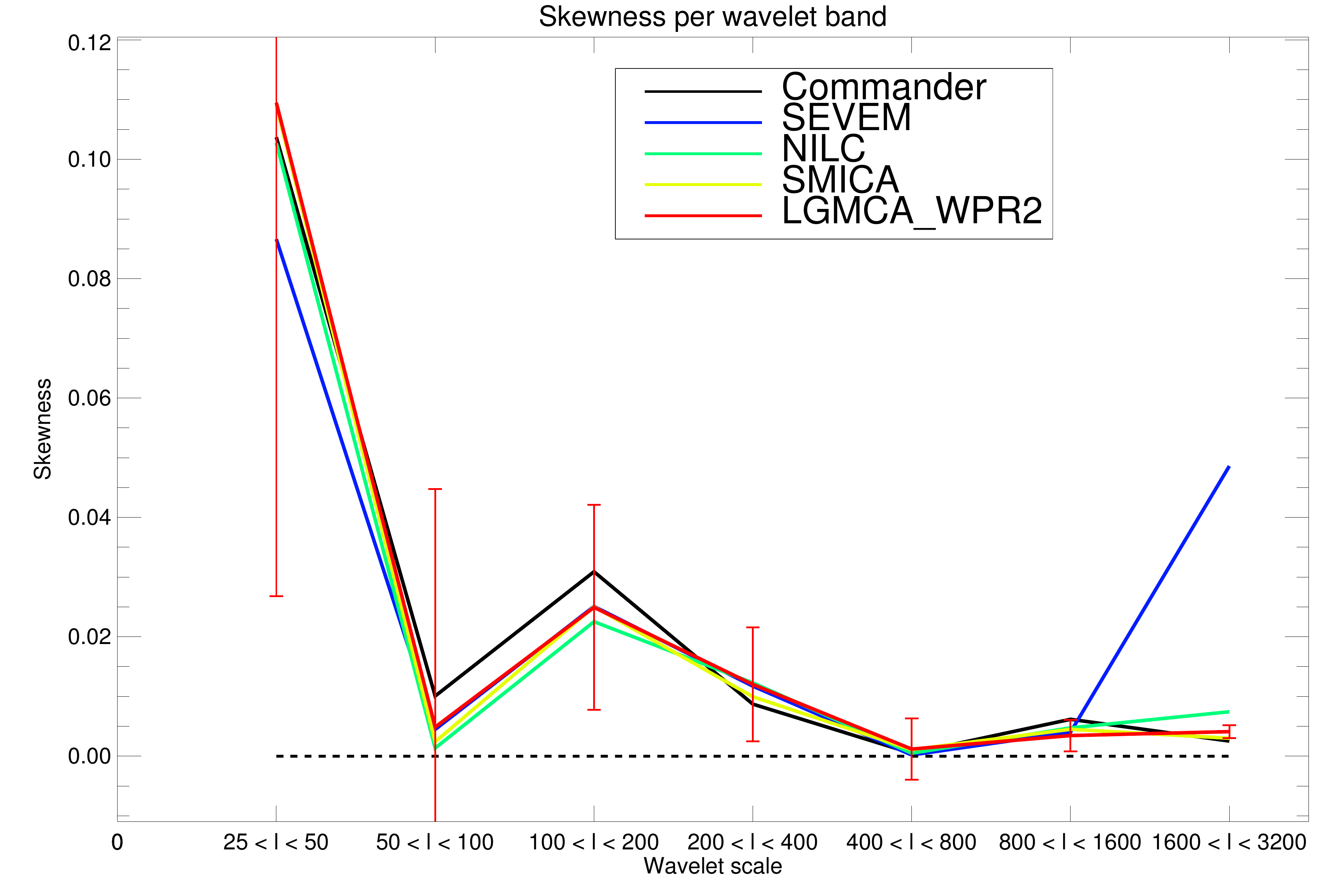}
\hspace{0.5cm}
\includegraphics [scale=0.25]{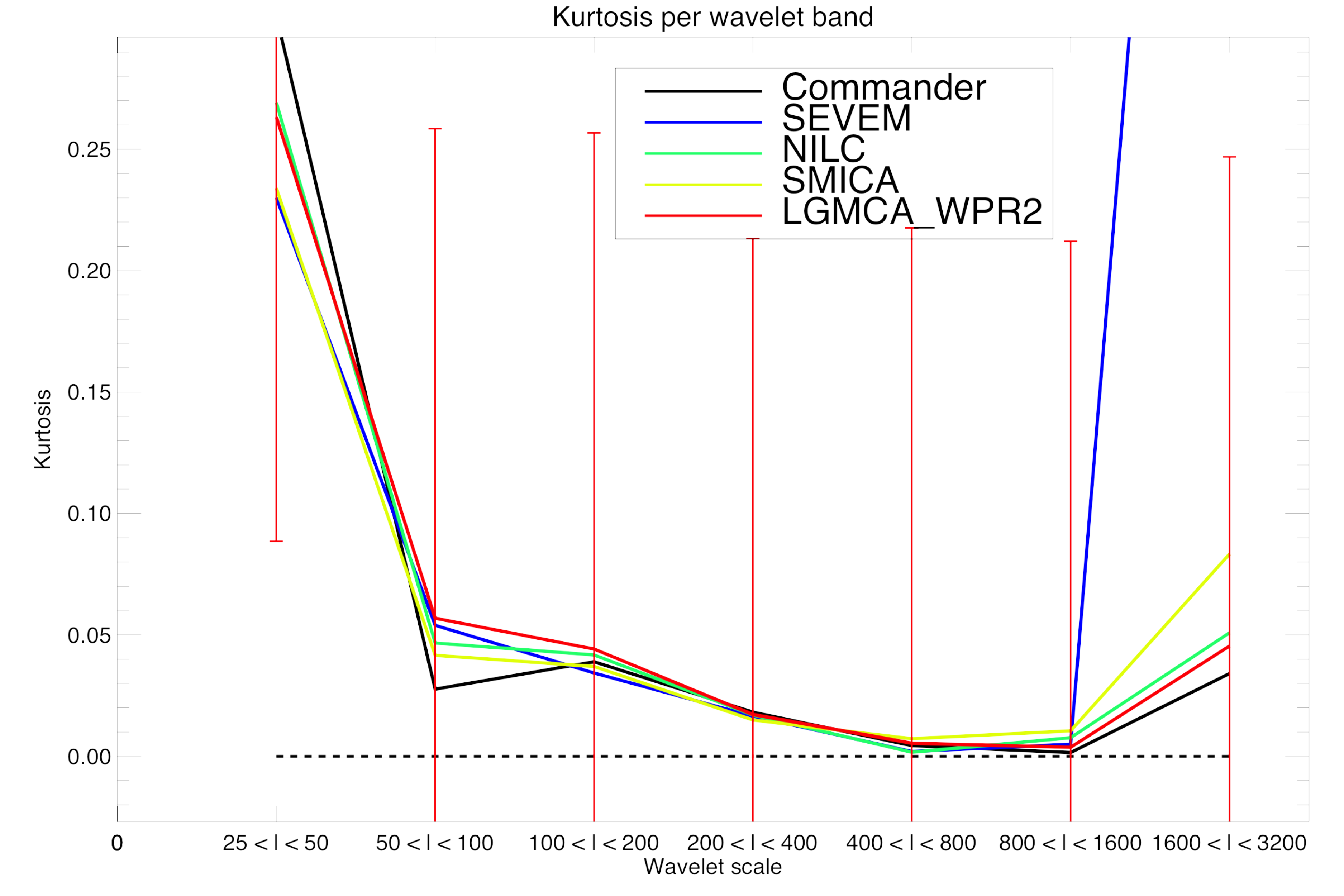}
}
}
\caption{High order statistics computed at various wavelet scales for the high resolution masks inside the $95\%$ analysis mask: Skewness (left), Kurtosis (top right). Error bars are set to $1\, sigma$, based on the noise level of the L-GMCA map}
\label{fig_HOS1}
\end{figure*}

\begin{figure*}[htb]
\centerline{
\hbox{
\includegraphics [scale=0.25]{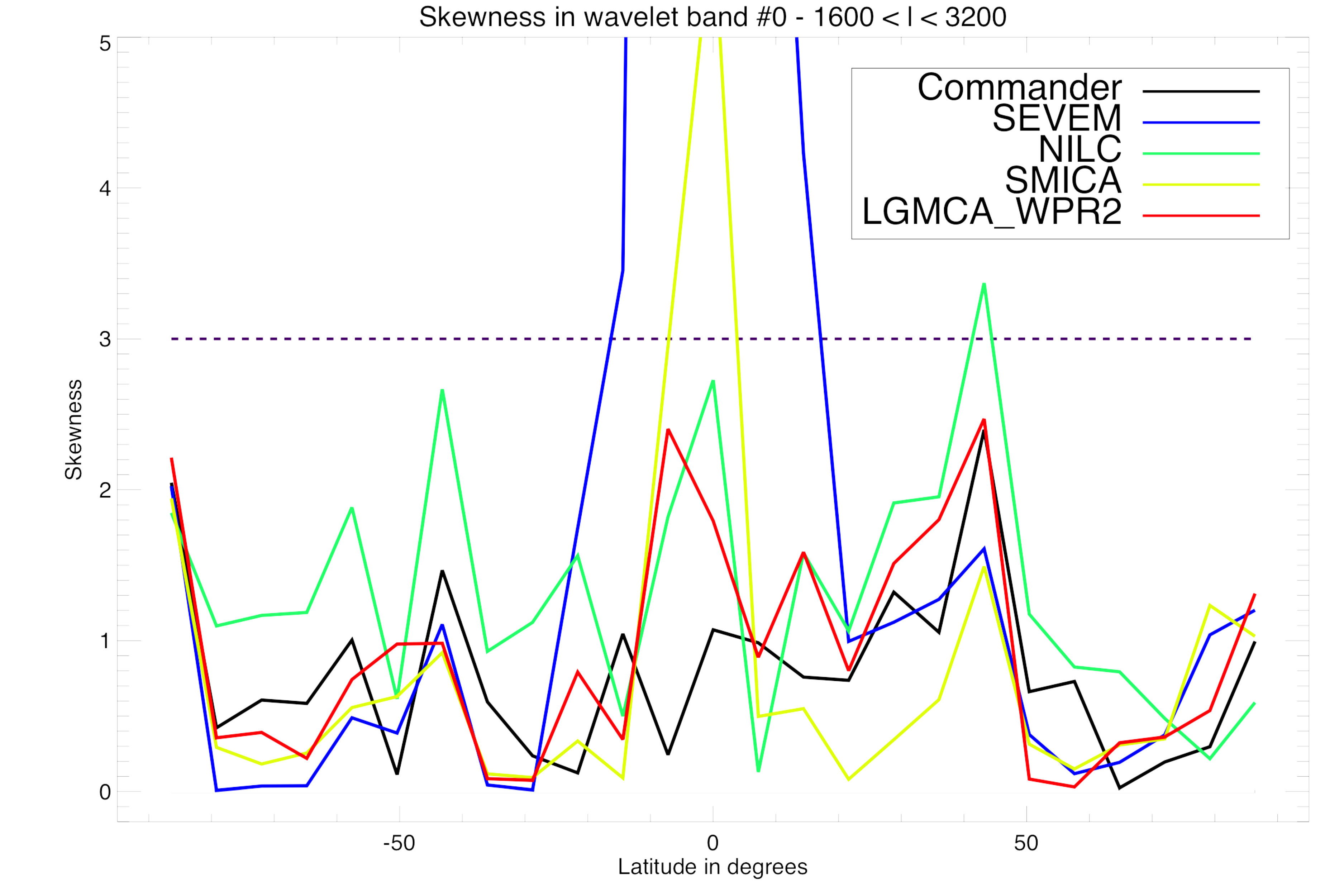}
\hspace{0.5cm}
\includegraphics [scale=0.25]{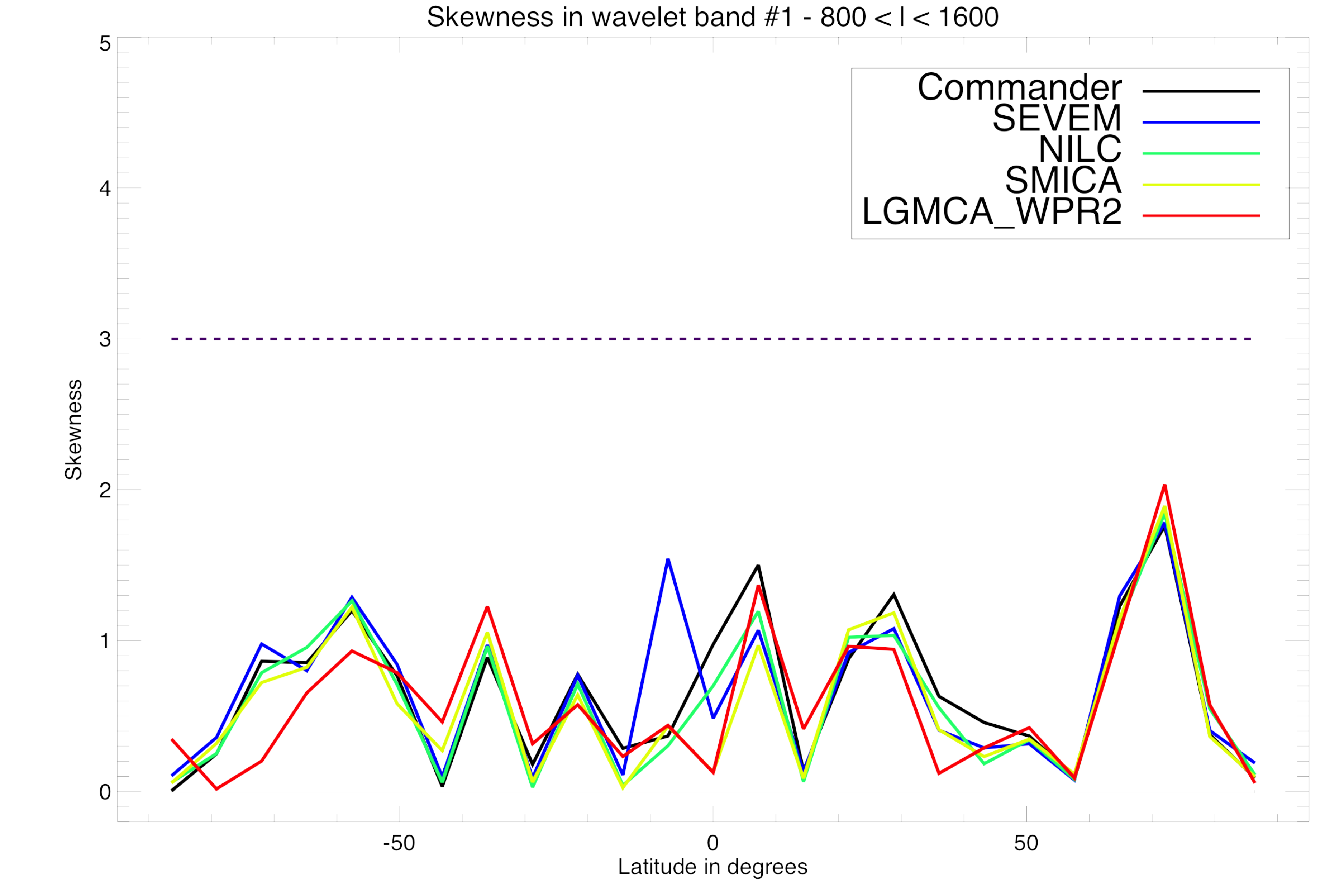}
}
}
\centerline{
\hbox{
\includegraphics [scale=0.25]{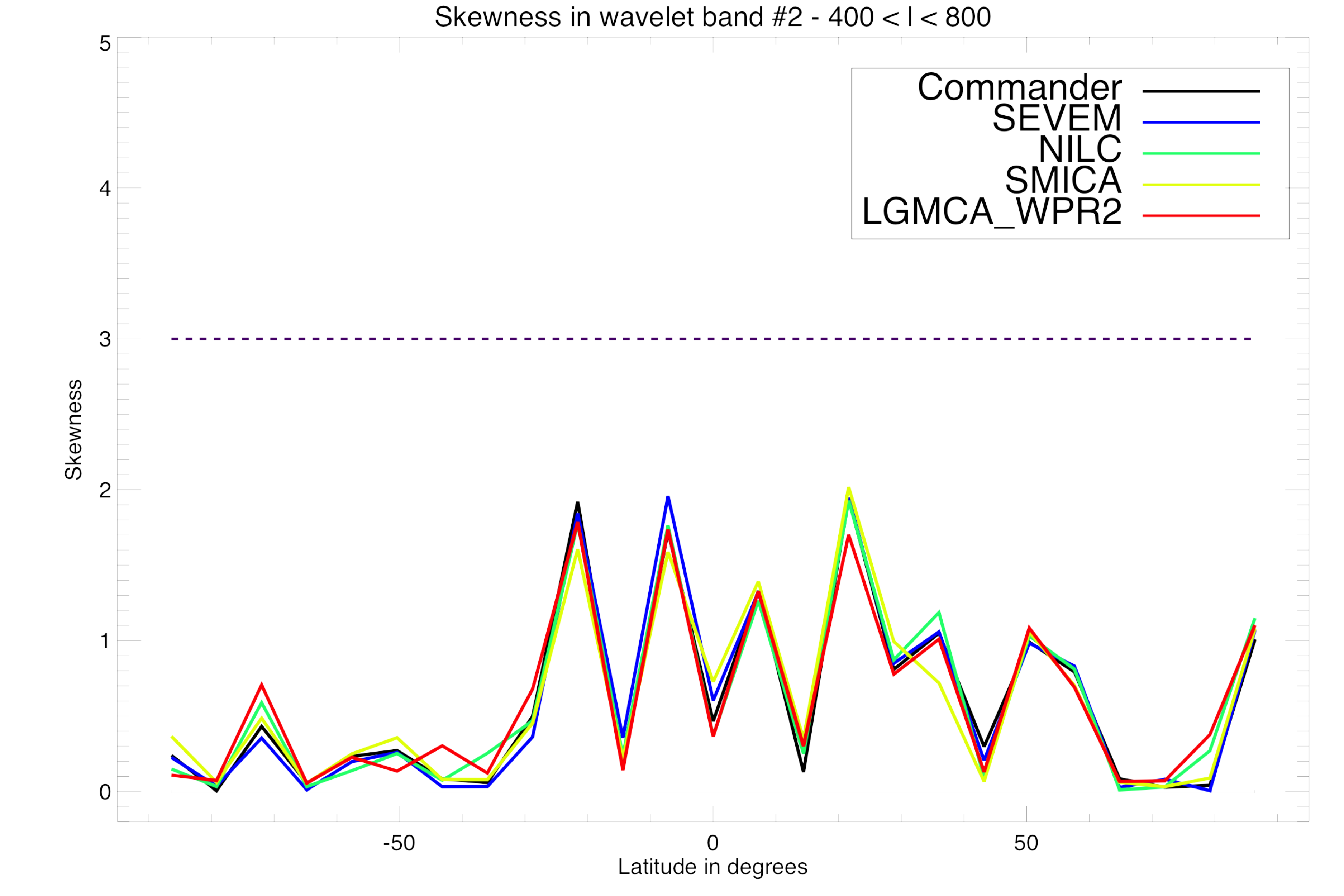}
\hspace{0.5cm}
\includegraphics [scale=0.25]{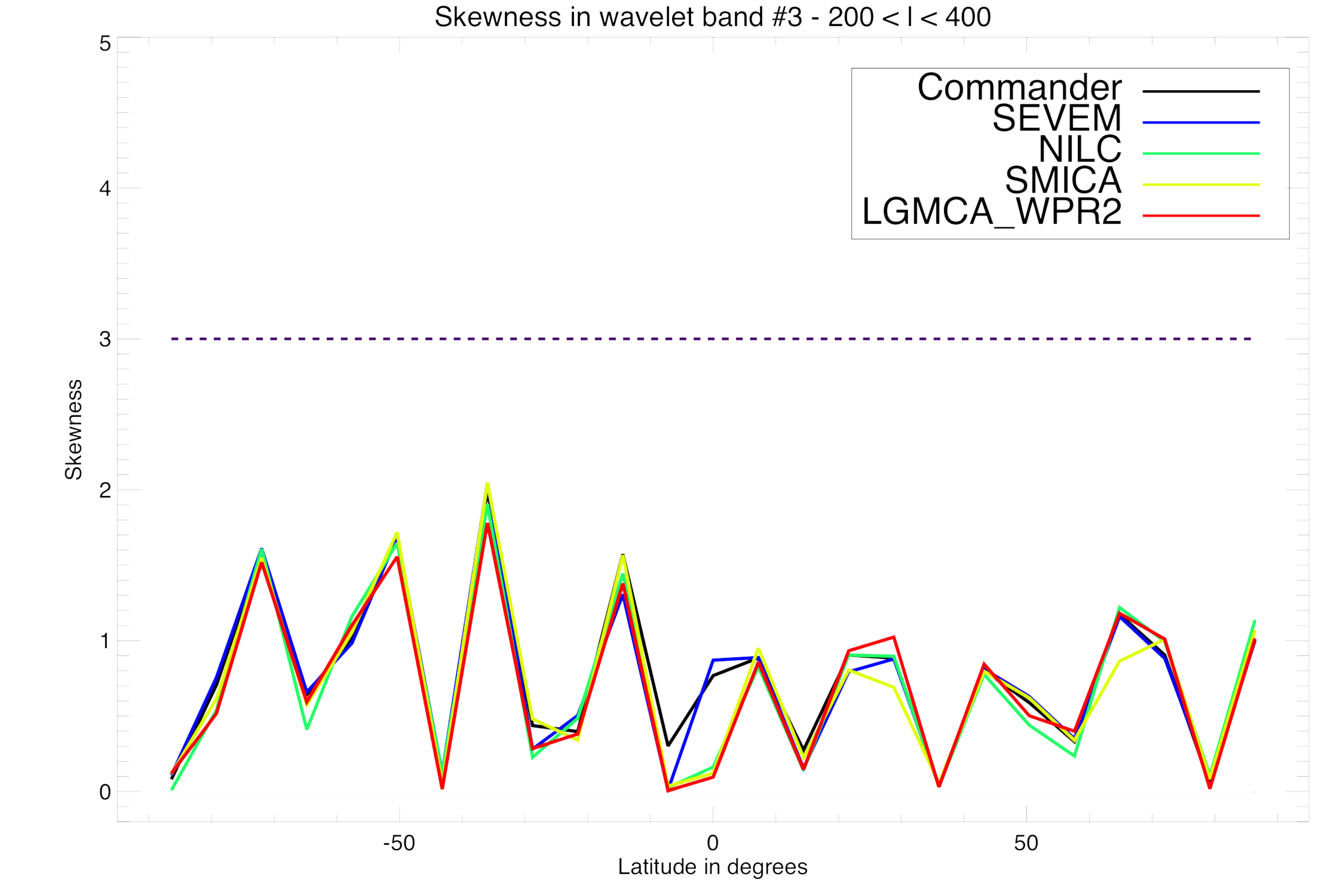}
}
}
\centerline{
\hbox{
\includegraphics [scale=0.25]{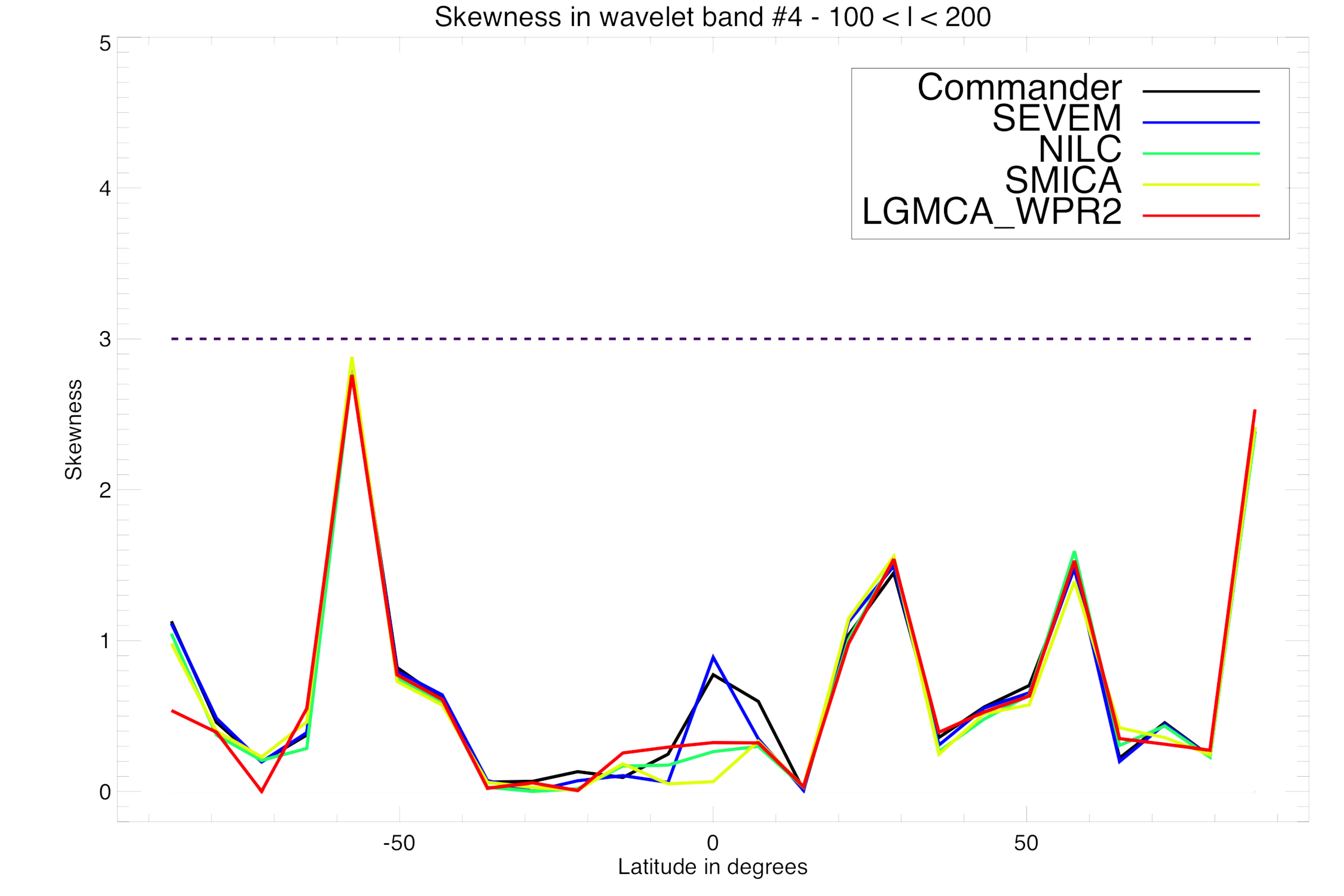}
\hspace{0.5cm}
\includegraphics [scale=0.25]{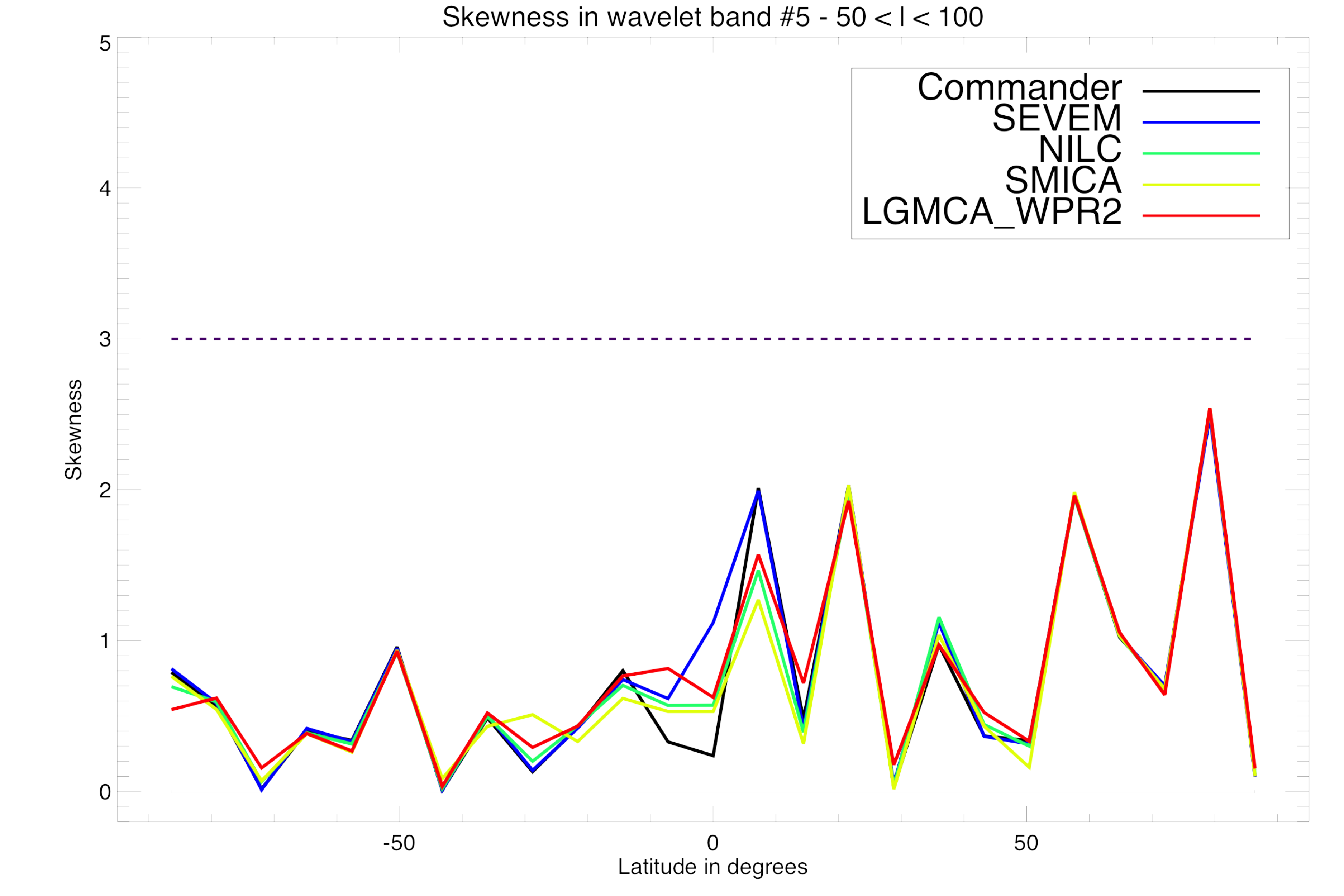}
}
}
\caption{Normalized skewness computed at various wavelet scales for the high resolution masks inside the $95\%$ analysis mask. The normalization is made with respect to the noise level of the L-GMCA map.}
\label{fig_HOS1_perlat}
\end{figure*}

\begin{figure*}[htb]
\centerline{
\hbox{
\includegraphics [scale=0.25]{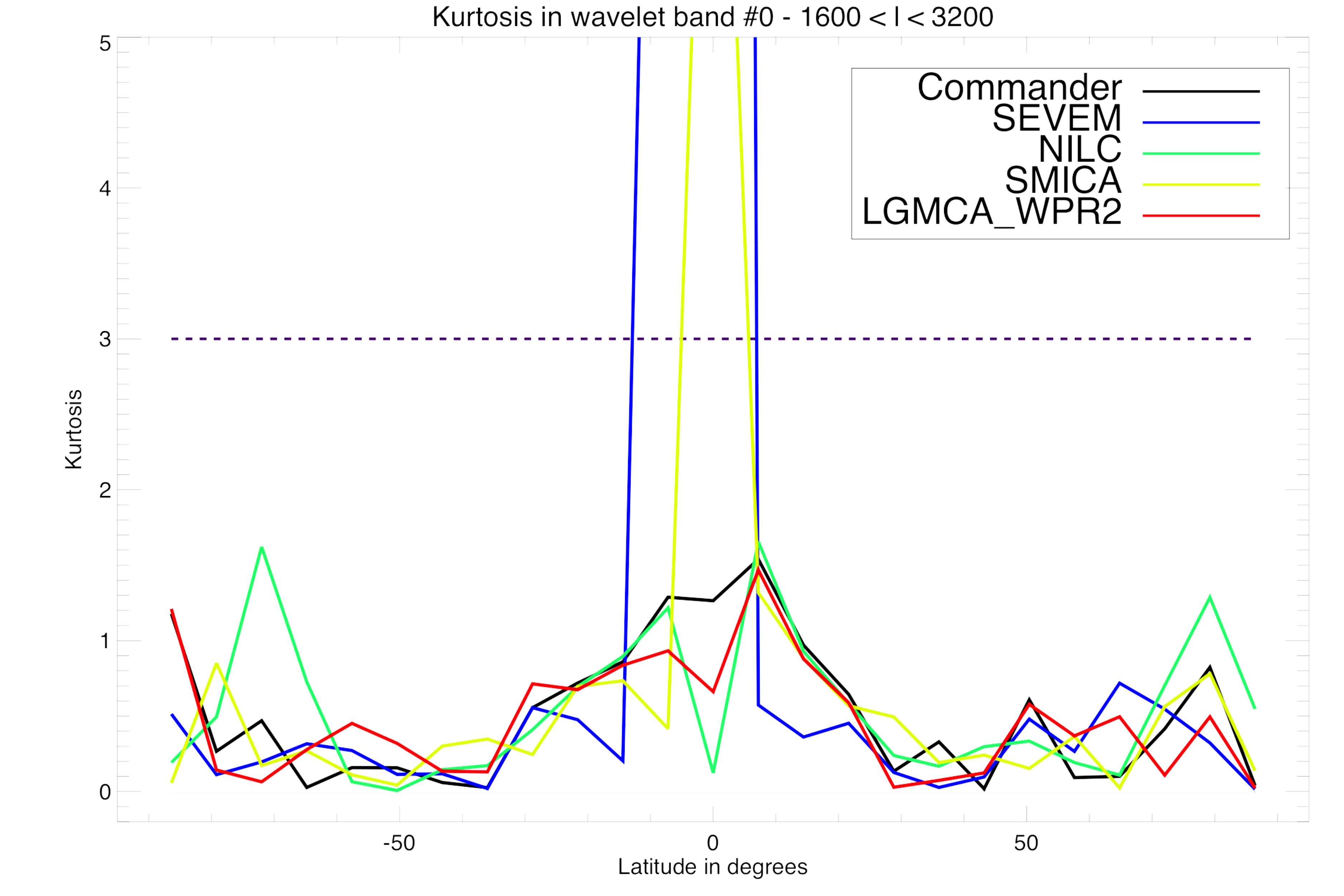}
\hspace{0.5cm}
\includegraphics [scale=0.25]{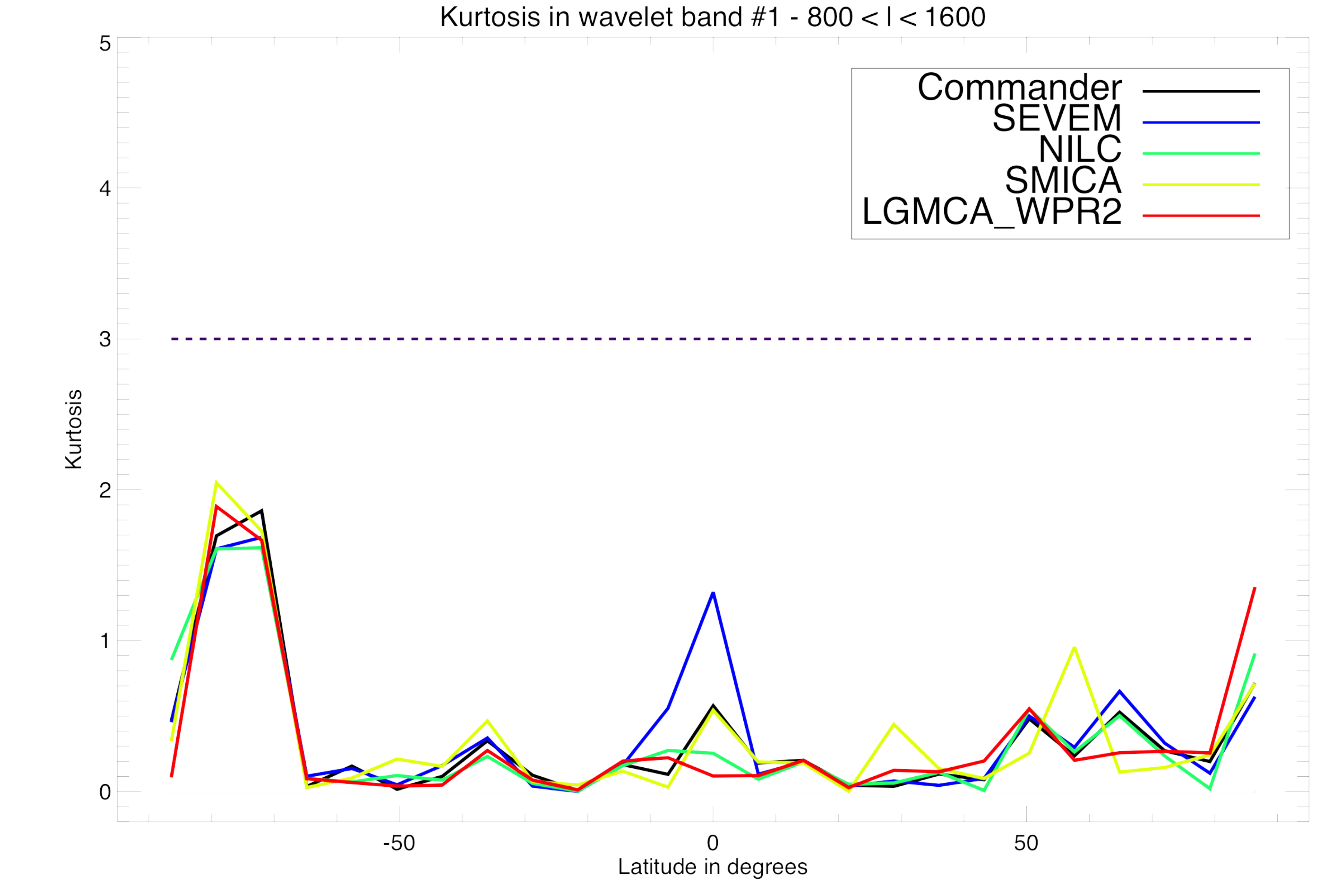}
}
}
\centerline{
\hbox{
\includegraphics [scale=0.25]{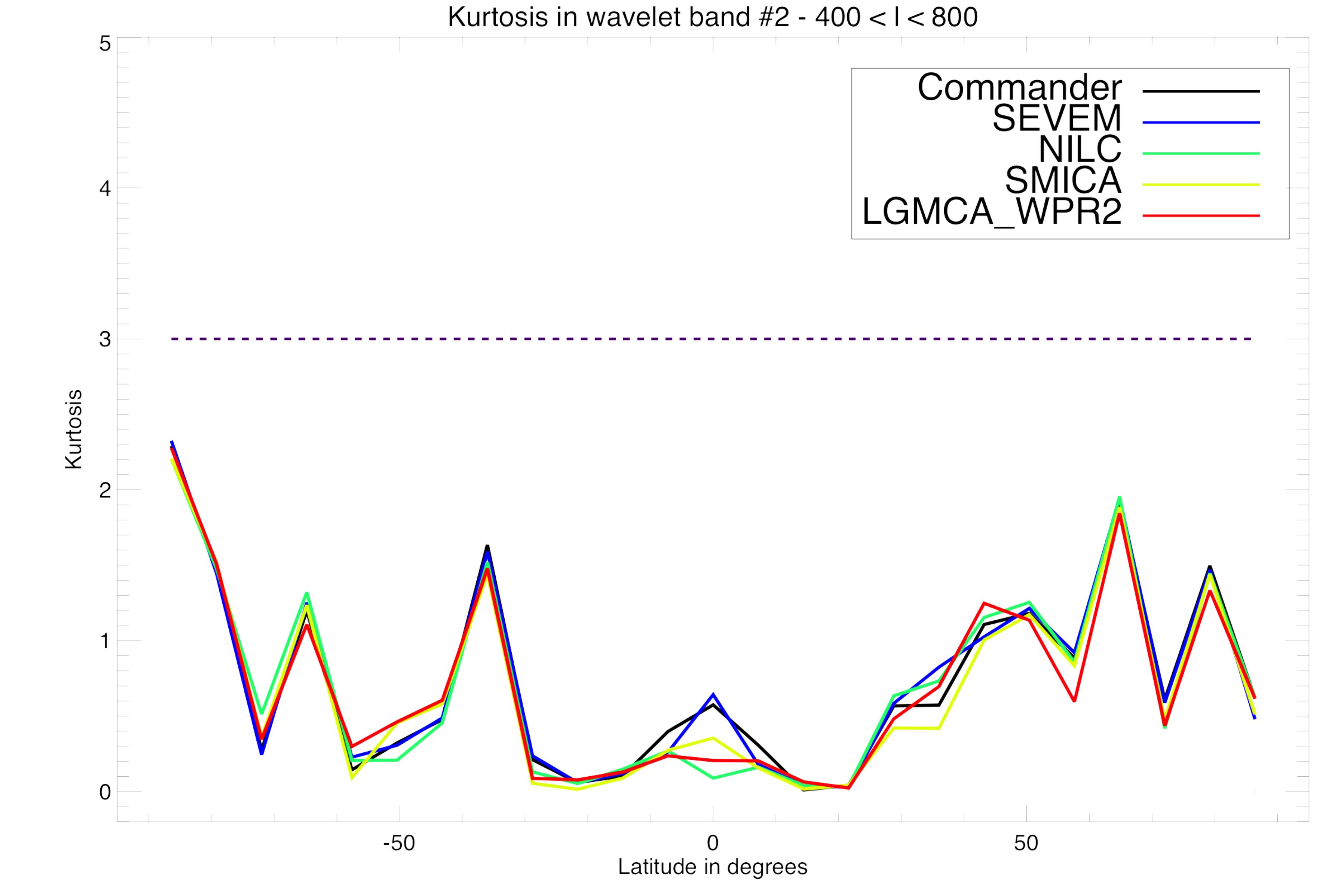}
\hspace{0.5cm}
\includegraphics [scale=0.25]{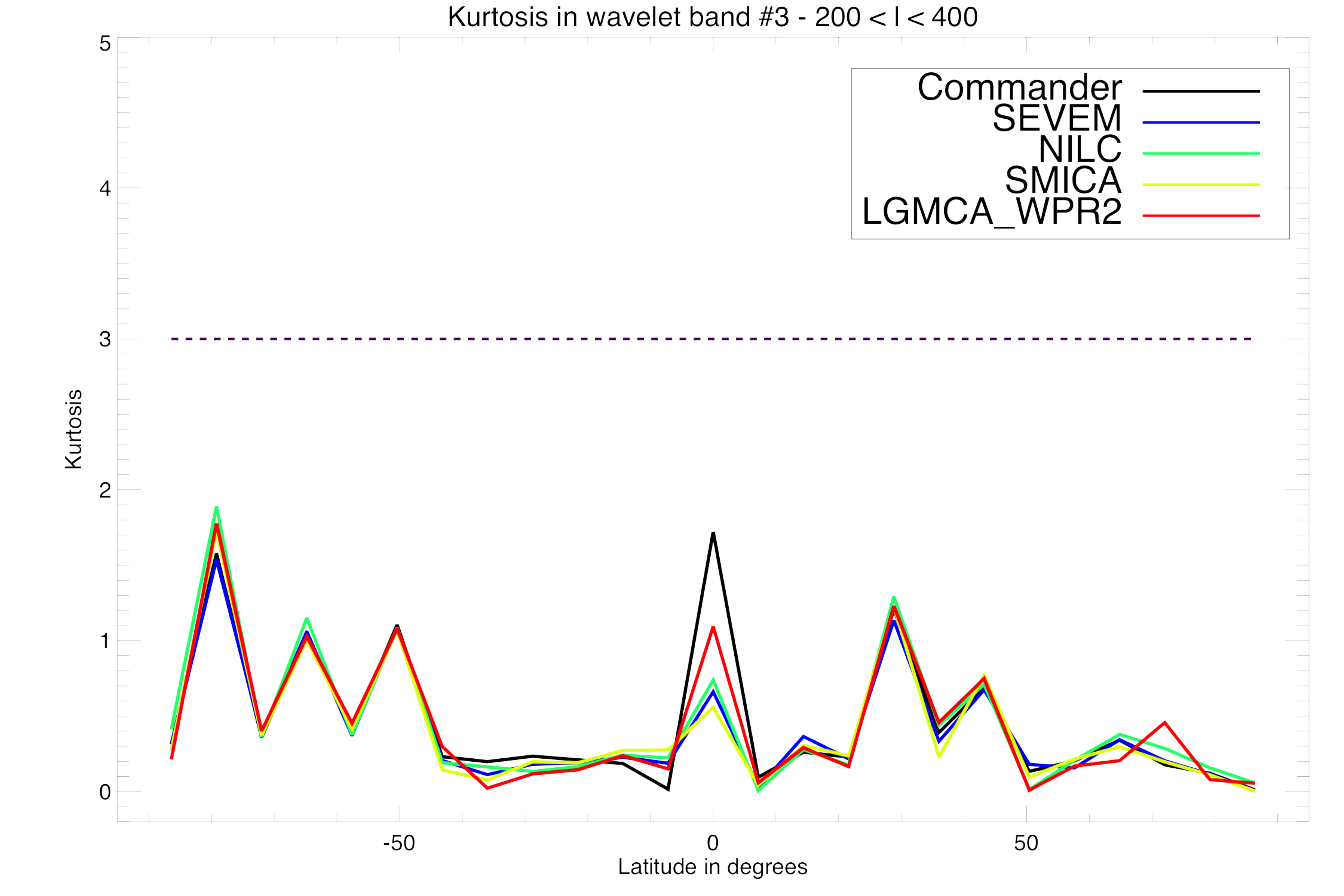}
}
}
\centerline{
\hbox{
\includegraphics [scale=0.25]{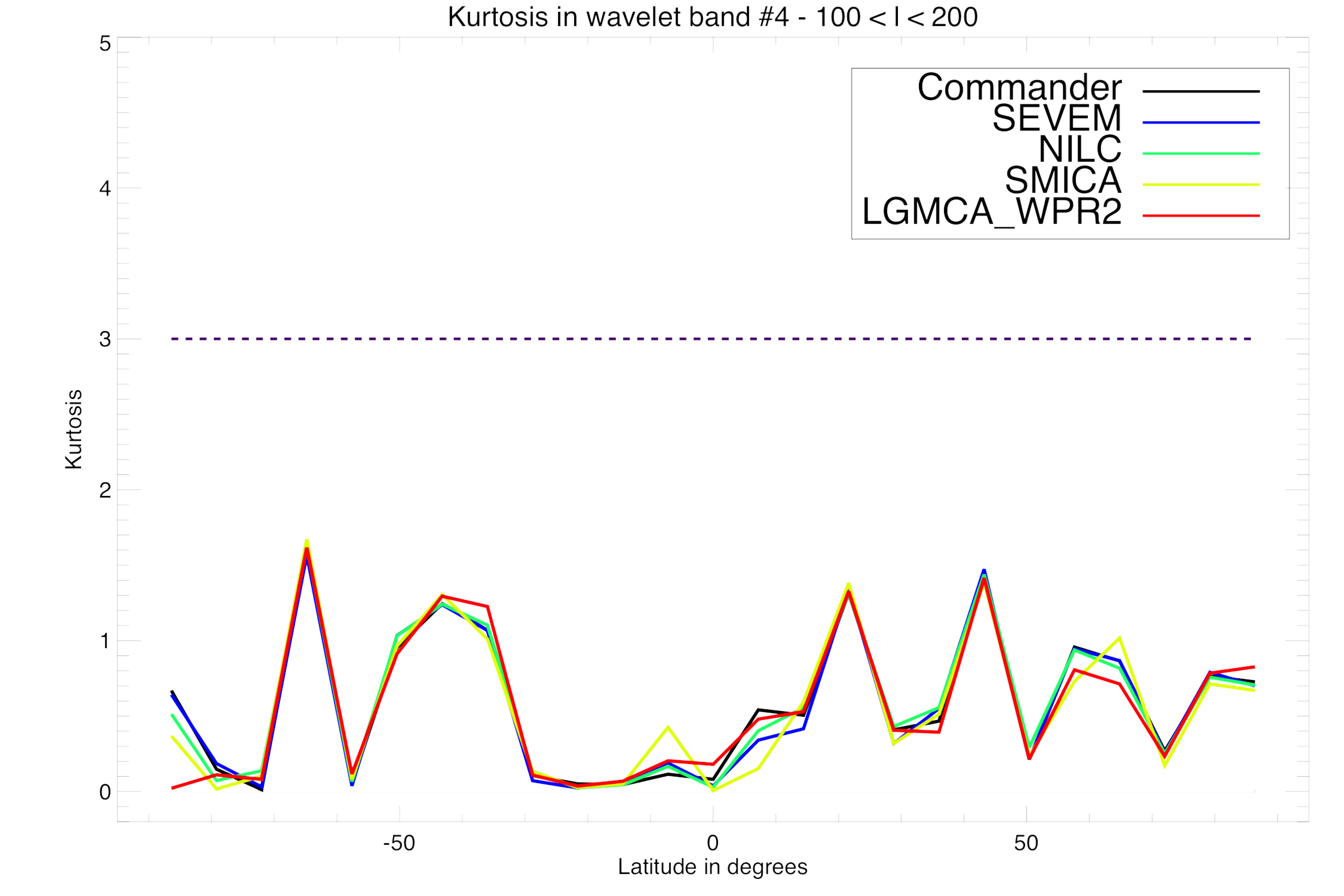}
\hspace{0.5cm}
\includegraphics [scale=0.25]{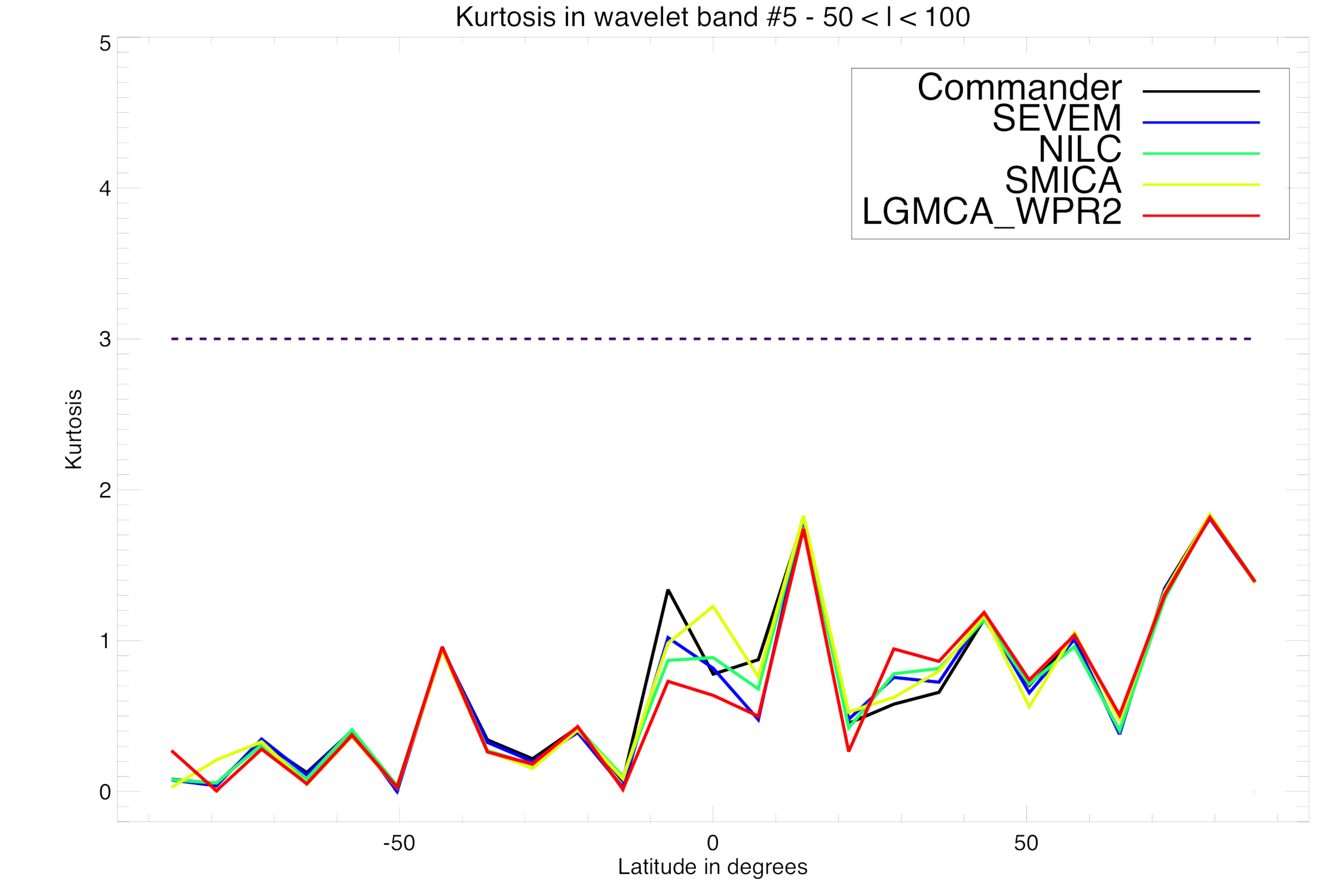}
}
}
\caption{Normalized kurtosis computed at various wavelet scales for the high resolution masks inside the $95\%$ analysis mask. The normalization is made with respect to the noise level of the L-GMCA map.}
\label{fig_HOS2_perlat}
\end{figure*}

Figure~\ref{fig_HOS1} shows the skewness (third-order cumulant) and the kurtosis (fourth-order cumulant) of the estimated CMB maps for the $7$ first wavelet scales. These values are computed using the same sky coverage defined by the combined SMICA mask of inpainted pixels. The error bars have been derived from $100$ Monte-Carlo simulations. Each Monte-Carlo simulation is composed of an independent CMB Gaussian realization, which we generated according to the Planck PR2 best-fit power spectrum. Independent noise realization have been computed based on the WPR2 L-GMCA map. With the exception of the CMB map reconstructed with SEVEM, we observe no strong departures from Gaussianity on the first seven scales. None of these maps exhibit non-Gaussianity for medium and large-scales ({\it i.e.} for wavelet scales corresponding to $\ell < 1600$). A refined characterization of the NG consists in computing the same statistics in different bands of latitude per wavelet scale.
Figure~\ref{fig_HOS1_perlat} shows the normalized skewness on scales $0$ to $5$ and Figure~\ref{fig_HOS2_perlat} bottom shows the kurtosis on scales $0$ to $5$. Normalization has been performed with respect to the error derived from $100$ Monte-Carlo simulations of a combination of pure CMB and noise maps derived from L-GMCA.\\

From this evaluation, we can conclude that, with the exception of the first wavelet scale, all maps are compatible with the Gaussianity assumption at all galactic latitudes for scales corresponding to $\ell < 1600$. In the finest scale, for $\ell > 1600$, the SEVEM and SMICA CMB maps show statistically significant non-Gaussian residuals, which are located on the galactic center. These features are detected both with their skewness and kurtosis statistics in the top-left panel of Figures~\ref{fig_HOS1_perlat} and \ref{fig_HOS2_perlat}. As well, the NILC map exhibits a slight level of non-Gaussianity at a galactic latitude of about $+45^{\circ}$. 
Apart from these features, all the maps are compatible with Gaussianity at all scales for galactic latitudes larger than $20^{\circ}$. The Commander CMB map exhibits no non-Gaussian features; however a significant part ({\it i.e.} about $18\%$ at small scales) has been masked and filled in with a constrained Gaussian realization. The L-GMCA WPR2 map is therefore the only one that does not show any non-Gaussian statistics, even on the galactic plane, which did not require any inpainting or interpolation procedure. 

\end{appendix}

\end{document}